\documentclass[twocolumn,9pt]{extarticle}
\usepackage[margin=2cm]{geometry}
\usepackage{natbib}
\usepackage{amsmath}
\usepackage{amssymb}
\usepackage{fontawesome5}
\usepackage{tikz}
\usetikzlibrary{positioning}
\usepackage{authblk}
\setkeys{Gin}{width=\columnwidth}
\usepackage{algorithm}
\usepackage[noend]{algpseudocode}

\DeclareMathOperator{\expect}{E}

\title{Finding mixed-strategy equilibria of continuous-action games without gradients using randomized policy networks}

\author[1]{Carlos Martin}
\author[1,2,3,4]{Tuomas Sandholm}
\affil[1]{Computer Science Department, Carnegie Mellon University}
\affil[2]{Strategy Robot, Inc.}
\affil[3]{Optimized Markets, Inc.}
\affil[4]{Strategic Machine, Inc.}
\date{}

\begin{document}

\maketitle

\begin{abstract}
We study the problem of computing an approximate Nash equilibrium of continuous-action game without access to gradients. Such game access is common in reinforcement learning settings, where the environment is typically treated as a black box. To tackle this problem, we apply zeroth-order optimization techniques that combine smoothed gradient estimators with equilibrium-finding dynamics.
We model players' strategies using artificial neural networks. In particular, we use randomized policy networks to model mixed strategies. These take noise in addition to an observation as input and can flexibly represent arbitrary observation-dependent, continuous-action distributions. Being able to model such mixed strategies is crucial for tackling continuous-action games that lack pure-strategy equilibria.
We evaluate the performance of our method using an approximation of the Nash convergence metric from game theory, which measures how much players can benefit from unilaterally changing their strategy.
We apply our method to continuous Colonel Blotto games, single-item and multi-item auctions, and a visibility game.
The experiments show that our method can quickly find high-quality approximate equilibria.
Furthermore, they show that the dimensionality of the input noise is crucial for performance.
To our knowledge, this paper is the first to solve general continuous-action games with unrestricted mixed strategies and without any gradient information.
\end{abstract}

\section{Introduction}

Most work on computing equilibria of games has focused on settings with finite, discrete action spaces. Yet many games involving space, time, money, \emph{etc.} actually have continuous action spaces. Examples include continuous resource allocation games~\citep{Ganzfried_2021}, security games in continuous spaces~\citep{Kamra_2017, Kamra_2018, Kamra_2019}, network games~\citep{Ghosh_2019}, simulations of military scenarios and wargaming~\citep{Marchesi20:Learning}, and video games~\citep{Berner19:Dota,Vinyals19:Grandmaster}. 
Furthermore, even if the action space is discrete, it may be fine-grained enough to treat as continuous for computational efficiency purposes~\citep{Borel38:Traite,Chen06:Mathematics,Ganzfried_2010}.

The typical approach to computing an equilibrium of a game with continuous action spaces involves discretizing the action space. That entails loss in solution quality~\citep{Kroer15:Discretization}. Also, it does not scale well; for one, in multidimensional action spaces it entails a combinatorial explosion of discretized points (exponential in the number of dimensions). Therefore, other approaches are called for. Furthermore, in many applications, explicit gradient information about the game is not available.

This paper is, to our knowledge, the first to solve general continuous-action games with unrestricted mixed strategies and without any gradient information.
We start by introducing some background needed to formulate the problem. We then describe related research that tackles the problem of computing approximate equilibria of such games. Next, we describe our method and its components, including smoothed gradient estimators, equilibrium-finding dynamics, and representation of mixed strategies\footnote{We use the terms \textit{policy} and \textit{strategy} interchangeably. The former is common in reinforcement learning, the latter in game theory.} using randomized policy networks. We then describe the various games that we use in the experiments. After that, we present our experimental results and discussion. Finally, we present our conclusion and suggest directions for future research.

\section{Problem description}

First, we introduce some notation: \(\triangle X\) is the set of probability measures on \(X\), \(\mathcal{U}(X)\) is the uniform probability measure on \(X\), and \([\cdot]\) is an Iverson bracket, which is 1 if its argument is true and 0 otherwise.

A \emph{strategic-form game} is a tuple \((I, S, u)\) where \(I\) is a set of players, \(S_i\) a set of strategies for player \(i\), and \(u_i : \prod_{j : I} S_j \to \mathbb{R}\) a utility function for player \(i\). A strategy profile \(s : \prod_{i : I} S_i\) maps players to strategies and \(s_{-i}\) denotes \(s\) excluding \(i\)'s strategy.
Player \(i\)'s best response utility \(b_i(s_{-i}) = \sup_{s_i : S_i} u_i(s)\) is the highest utility they can attain given the other players' strategies. Their utility gap \(g_i(s) = b_i(s_{-i}) - u_i(s)\) is the highest utility they can gain from unilaterally changing their strategy, and \(s\) is an \(\varepsilon\)-equilibrium iff \(\sup_{i : I} g_i(s) \leq \varepsilon\). A 0-equilibrium is called a Nash equilibrium.
A common measure of closeness to Nash equilibrium is \emph{NashConv}, defined as \(\bar{g} = \int_{i \sim \mu} g_i\), where \(\mu\) is some measure on \(I\). Typically, \(I\) is finite and \(\mu\) is the counting measure, making \(\bar{g}\) a finite sum of utility gaps.
However, some games may have infinitely many players, such as mean field games.
A game is zero-sum iff \(\int_{i \sim \mu} u_i = 0\), which makes \(\bar{g} = \int_{i \sim \mu} b_i\).
In a two-player zero-sum game, \(\bar{g}\) reduces to the so-called ``duality gap''~\citep{Grnarova_2019}: \(\bar{g}(s) = \sup_{s_1'} u(s_1', s_2) - \inf_{s_2'} u(s_1, s_2')\).

In many games, the \(S_i\) are infinite. The following theorems apply to such games. 
If for all \(i\), \(S_i\) is nonempty and compact, and \(u_i\) is continuous in \(s\), a mixed strategy Nash equilibrium exists~\citep{Glicksberg52:Further}.
If for all \(i\), \(S_i\) is nonempty, compact, and convex, and \(u_i\) is continuous in \(s\) and quasi-concave in \(s_i\), a pure strategy Nash equilibrium exists~\citep[p. 34]{Fudenberg91:Gamea}.
Other results include the existence of a mixed strategy Nash equilibrium for games with discontinuous utilities under some mild semicontinuity conditions on the utility functions~\citep{Dasgupta86:Existence}, and the uniqueness of a pure Nash equilibrium for continuous games under diagonal strict concavity assumptions~\citep{Rosen_1965}.

A \emph{Bayesian game} is a game of incomplete information, that is, a game in which players have only partial information about the game and other players. Formally, it is a tuple \((I, \Omega, \mu, O, \tau, A, r)\) where \(I\) is a set of players, \(\Omega\) a set of states, \(\mu : \triangle \Omega\) a distribution over states, \(O_i\) a set of observations for \(i\), \(\tau_i : \Omega \to O_i\) an observation function for \(i\), \(A_i\) a set of actions for \(i\), and \(r_i : \Omega \times \prod_{j : I} A_j \to \mathbb{R}\) a payoff function for \(i\). A strategy for player \(i\) is a function \(s_i : O_i \to \triangle A_i\). Given strategy profile \(s\), player \(i\)'s expected payoff is \(u_i(s) = \expect_{\omega \sim \mu} \expect_{a_j \sim s_j(\tau_j(\omega)), \forall j : I} r_i(\omega, a)\) and their best response payoff \(b_i(s_{-i}) = \sup_{s_i} u_i(s)\) is
\begin{align}
    \sup_{s_i} \expect_{\omega} \expect_a r_i(\omega, a)
    &= \expect_{o_i} \sup_{s_i} \expect_{\omega | o_i} \expect_a r_i(\omega, a) \\
    &= \expect_{o_i} \sup_{a_i} \expect_{\omega | o_i} \expect_{a_{-i}} r_i(\omega, a)
    \label{eq:br}
\end{align}
where \(\omega | o_i\) conditions \(\omega\) on player \(i\)'s observation being \(o_i\).

\section{Related research}

\citet{McMahan_2003} introduce the double oracle algorithm for normal-form games and prove convergence.~\citet{Adam_2021} extended it to two-player zero-sum continuous games.~\citet{Kroupa_2021} extend it to \(n\)-player continuous games.
Their algorithm maintains finite strategy sets for each player and iteratively extends them with best responses to an equilibrium of the induced finite sub-game.
It ``converges fast when the dimension of strategy spaces is small, and the generated subgames are not large.''
For example, in the two-player zero-sum case: ``The best responses were computed by selecting the best point of a uniform discretization for the one-dimensional problems and by using a mixed-integer linear programming reformulation for the Colonel Blotto games.'' This approach does not scale to high-dimensional games with general payoffs where best-response computation is difficult. Moreover, if the game is stochastic, estimating the finite subgame can be difficult and require many samples. Furthermore, this approach does not learn observation-dependent strategies that generalize across observations.

\citet{Ganzfried_2021} introduce an algorithm for approximating equilibria in continuous games called  ``redundant fictitious play'' and apply it to a continuous Colonel Blotto game.
\citet{Kamra_2019} present DeepFP, an approximate extension of fictitious play~\citep{Brown51:Iterative, Berger_2007} to continuous action spaces.
They demonstrate stable convergence to equilibrium on several classic games and a large forest security domain.
DeepFP represents players' approximate best responses via generative neural networks.
The authors state that such models cannot be trained directly in the absence of gradients, and thus employ a game-model network that is a differentiable approximation of the game's payoff function, training these networks end-to-end in a model-based learning regime.
Our approach shows, however, that these generative models \emph{can} be trained directly.

\citet{Li_2021} extend the double oracle approach to \(n\)-player general-sum continuous Bayesian games. They represent agents as neural networks and optimize them using \emph{natural evolution strategies (NES)}~\citep{Wierstra_2008, Wierstra_2014}. To approximate a pure-strategy equilibrium, they formulate the problem as a bi-level optimization and employ NES to implement both inner-loop best response optimization and outer-loop regret minimization.
\citet{Bichler_2021} represent strategies as neural networks and apply simultaneous gradients to provably learn local equilibria. They focus on symmetric auction models, assuming symmetric prior distributions and symmetric equilibrium bidding strategies.~\citet{Bichler_2022} extend that to asymmetric auctions, where one needs to train multiple neural networks.
Both papers restrict their attention to pure strategies.
~\citet{Fichtl_2022} compute distributional strategies~\citep{Milgrom85:Distributional} (a form of mixed strategies for Bayesian game) on a discretized version of the game via online convex optimization, specifically \emph{simultaneous online dual averaging}, and show that the equilibrium of the discretized game approximates an equilibrium in the continuous game. That is, they discretize the type and action spaces and implement gradient dynamics in the discretized version of the game without using neural networks. 
In contrast, our approach does not use discretization, which can work well for small games but does not scale to high-dimensional observation and action spaces.

A \emph{generative adversarial network (GAN)}~\citep{Goodfellow_2014} consists of a generator that learns to generate fake data and a discriminator that learns to distinguish it from real data.~\citet{Grnarova_2019} propose using an approximation of the game-theoretic \emph{duality gap} as a performance measure for GANs.~\citet{Grnarova_2021} propose using this measure as the training objective itself, proving some convergence guarantees.
\citet{Lockhart_2019} present exploitability descent, which computes approximate equilibria in two-player zero-sum extensive-form games by direct strategy optimization against worst-case opponents. They prove that the exploitability of a player's strategy converges asymptotically to zero. Hence, if both players employ this optimization, the strategy profile converges to an equilibrium. Unlike extensive-form fictitious play~\citep{Heinrich_2015} and counterfactual regret minimization~\citep{Zinkevich07:Regret}, their result pertains to last-iterate rather than average-iterate convergence.
\citep{Timbers_2022} introduce approximate exploitability, which uses an approximate best response computed through search and reinforcement learning.
This is a generalization of local best response, a domain-specific evaluation metric used in poker~\citep{Lisy17:Equilibrium}.

\citet{Gemp_2022} propose an approach called \emph{average deviation incentive descent with adaptive sampling} that iteratively improves an approximation to a Nash equilibrium through joint play by tracing a homotopy that defines a continuum of equilibria for the game regularized with decaying levels of entropy. To encourage iterates to remain near this path, they minimize average deviation incentive via stochastic gradient descent.

\citet{Ganzfried10:Computing, Ganzfried_2010} present a procedure for solving large imperfect-information games by solving an infinite approximation of the original game and mapping the equilibrium to a strategy profile in the original game.
Perhaps counterintuitively, the infinite approximation can often be solved much more easily than the finite game.
The algorithm exploits some qualitative model of equilibrium structure as an additional input to find an equilibrium in continuous games.

\section{Game-solving technique}
We now describe the components of our game-solving technique.
\subsection{Gradient estimation}

Consider the problem of maximizing \(f : \mathbb{R}^d \to \mathbb{R}\) with access to its values but not derivatives. This setting is called \emph{zeroth-order optimization}. One approach to this problem is to compute estimates of the gradient \(g(x) \approx \nabla f(x)\) and apply gradient-based optimization. The gradient could be estimated via finite differences as \(g(x)_i = \tfrac{1}{\sigma} (f(x + \sigma e_i) - f(x))\) for all \(i \in [d]\), where \(e_i\) is the \(i\)th standard basis vector and \(\sigma\) is a small number. However, the number of queries needed scales linearly with the number of dimensions \(d\). Another approach is to evaluate the function at \emph{randomly-sampled} points and estimate the gradient as a sum of estimates of directional derivatives along random directions~\citep{Duchi_2015, Nesterov_2017, Shamir_2017, Berahas_2022}. These methods compute an unbiased estimator of the gradient of a \emph{smoothed} version of \(f\) induced by stochastically perturbing the input under some distribution \(\mu_1\) and taking the expectation~\citep{Duchi12:Randomized}. Specifically, for distributions \(\mu_1\) and \(\mu_2\),
\(
\nabla_x \expect_{u \sim \mu_1} f(x + \sigma u) = \tfrac{1}{\sigma} \expect_{u \sim \mu_2} f(x + \sigma u) u
\).
Gaussian smoothing uses \(\mu_1 = \mu_2 = \mathcal{N}(0, I_d)\). Ball smoothing uses \(\mu_1 = \mathcal{U}(\sqrt{d} \mathbb{B}_d), \mu_2 = \mathcal{U}(\sqrt{d} \mathbb{S}_d)\), where \(\mathbb{B}_d\) and \(\mathbb{S}_d\) are the \(d\)-dimensional unit ball and sphere.
These yield instances of a class of black box optimization algorithms called \emph{evolution strategies}~\citep{Rechenberg_1973, Schwefel_1977, Rechenberg_1978}, which maintain and evolve a population of parameter vectors. Specifically, they yield instances of \emph{natural evolution strategies}~\citep{Wierstra_2008, Wierstra_2014, Sun_2009}, which represent the population as a distribution over parameters and maximize its average objective value using the score function estimator. For example, Gaussian smoothing has been applied to single-agent reinforcement learning and obtains competitive results on standard benchmarks~\citep{Salimans_2017}.

To estimate the smoothed gradient, various \emph{stencils} can be used. These have the form \(\tfrac{1}{\sigma N} \sum_{i=1}^N a_i u_i\) where \(u_i \sim \mu_2\) independently and \(a_i\) is \(f(x + \sigma u_i)\), \(f(x + \sigma u_i) - f(x)\), and \(\tfrac{1}{2}(f(x + \sigma u_i) - f(x - \sigma u_i))\) for the single-point, forward-difference, and central-difference stencils, respectively. The single-point stencil has a large variance that diverges to infinity as \(\sigma\) approaches 0, so the latter two are typically used in practice~\citep{Berahas_2022}.
A related method is \emph{simultaneous-perturbation stochastic approximation (SPSA)}~\citep{Spall_1992}, which perturbs each coordinate with Rademacher variates \(u \sim \mathcal{U}(\{-1, 1\}^d)\) and uses the central-difference stencil.
A Taylor expansion of \(f\) shows that this is a good estimate of the true gradient when \(\sigma\) is small.
\citet{Spall_1997} introduced a one-measurement variant of SPSA that uses the single-point stencil.
In our experiments, we use Gaussian smoothing with the central-difference stencil.

\subsection{Equilibrium-finding dynamics}
Several gradient-based algorithms exist for finding equilibria in continuous games, as described in the appendix. Their convergence is analyzed in various works, including~\citet{Balduzzi_2018, Letcher_2019, Mertikopoulos_2019, Grnarova_2019, Mazumdar_2019, Hsieh_2021}.
In the games we tested, simultaneous gradient ascent was sufficient to obtain convergence to equilibrium and the other dynamics did not yield further improvements.
\citet{Mertikopoulos_2019} analyze the conditions under which simultaneous gradient ascent converges to Nash equilibria. They prove that, if the game admits a pseudoconcave potential or if it is monotone, the players' actions converge to Nash equilibrium, no matter the level of uncertainty affecting the players' feedback.
\citet{Bichler_2021} write that most auctions in the literature assume symmetric bidders and symmetric equilibrium bid functions~\citep{Krishna02:Auction}.
This symmetry creates a potential game, and simultaneous gradient dynamics provably converge to a pure local Nash equilibria in finite-dimensional continuous potential games~\citep{Mazumdar_2020}.
Thus in any symmetric and smooth auction game, symmetric gradient ascent with appropriate (square-summable but not summable) step sizes almost surely converges to a local ex-ante approximate Bayes-Nash equilibrium~\citep[Proposition 1]{Bichler_2021}. These results apply to most of our experiments, except for the asymmetric-information auction.

\subsection{Distributed training algorithm}

Our algorithm for training strategy profiles can also be efficiently distributed, as we now describe. The pseudocode is given as Algorithm \ref{alg:algorithm}.
On any iteration, there is a set of available workers \(\mathcal{J}\). Each worker is assigned the task of computing a pseudogradient for a particular player. The vector \(\{a_j\}_{j \in \mathcal{J}}\) contains the assignment of a player for each worker.
Each worker's pseudorandom number generator (PRNG) is initialized with the same fixed seed.
On any iteration, one of the workers is the \emph{coordinator}.
Initially, or when the current coordinator goes offline, the workers choose a coordinator by running a leader election algorithm.

On each iteration, each worker evaluates the utility function (generally the most expensive operation and bottleneck for training) twice to compute the finite difference required for the pseudogradient. It then sends this computed finite difference (a single scalar) to the coordinator.
The coordinator then sends the vector of these scalars to every worker, ensuring that all workers see each other's scalars.
Thus the information that needs to be passed between workers is minimal. This greatly reduces the cross-worker bandwidth required by the algorithm compared to schemes that pass parameters or gradients between workers, which can be very expensive for large models.

\algnewcommand\algorithmicinput{\textbf{Input:}}
\algnewcommand\Input{\item[\algorithmicinput]}
\begin{algorithm}
\caption{Distributed multiagent pseudogradient ascent}
\label{alg:algorithm}
\begin{algorithmic}
\State The following algorithm runs on each worker:
\Input \(\mathcal{I}\) is the set of players
\Input \(u\) is the utility function
\State initialize PRNG state with fixed seed
\State \(\mathbf{x} \gets\) initial strategy profile
\For{\(i \in \mathcal{I}\)}
\State \(S_i \gets \text{init}(\mathbf{x}_i)\) \Comment initial state of optimizer \(i\)
\EndFor
\Loop
\State \(\mathcal{J} \gets\) set of available workers
\For{\(j \in \mathcal{J}\)}
\State \(a_j \gets\) player \(\in \mathcal{I}\) \Comment can be set dynamically
\State \(\varepsilon_j \gets\) scale \(\in \mathbb{R}_{> 0}\) \Comment can be set dynamically
\State \(\mathbf{z}_j \sim N(\mathbf{0}, \mathbf{I}_{\dim \mathbf{x}_i})\) where \(i = a_j\)
\EndFor
\State \(j \gets \) own worker ID
\State \(\delta_j \gets \frac{u(\mathbf{x}_i + \varepsilon_j \mathbf{z}_j, \mathbf{x}_{-i})_i - u(\mathbf{x}_i - \varepsilon_j \mathbf{z}_j, \mathbf{x}_{-i})_i}{2 \varepsilon_j}\) where \(i = a_j\)
\State send \(\delta_j\) to coordinator
\State receive \(\delta\) from coordinator
\For{\(i \in \mathcal{I}\)}
\State \(\mathcal{K} \gets \{j \in \mathcal{J} \mid a_j = i\}\) \Comment workers assigned \(i\)
\State \(\mathbf{v}_i \gets \frac{1}{|\mathcal{K}|} \sum_{j \in \mathcal{K}} \delta_j \mathbf{z}_j\) \Comment \(i\)'s pseudogradient
\State \(S_i, \mathbf{x}_i \gets \text{step}(S_i, \mathbf{v}_i)\) \Comment step optimizer \(i\)
\EndFor
\EndLoop
\end{algorithmic}
\end{algorithm}

This massively parallelizes~\citep[Algorithm~1]{Bichler_2021} (``NPGA using ES gradients'').
Simultaneously, it generalizes~\citep[Algorithm~2]{Salimans_2017} (``Parallelized Evolution Strategies''), which also uses shared seeds, to the multiplayer setting, with separate gradient evaluations and optimizers for each player.
Furthermore, it allows for the possibility of setting the worker-player assignments \(a_j\) and perturbation noise scales \(\varepsilon_j\) dynamically over time, provided that this is done  consistently across workers (e.g., based on their common state variables).
Vanilla gradient descent, momentum gradient descent, optimistic gradient descent, or other optimization algorithms can be used. More information about the different optimization algorithms is in the appendix in the section on alternative equilibrium-finding dynamics.

The set of available workers can also change dynamically over time. If a worker leaves or joins the pool, the coordinator notifies all workers of its ID so they can remove it from, or add it to, their \(\mathcal{J}\) sets, respectively. The new worker is brought up to speed by passing it the current PRNG state, strategy profile parameters, and optimizer states (what state information is needed depends on the algorithm used, for example, whether momentum is used).

\subsection{Policy representation}

Another key design choice is how the players' strategies are modeled.~\citet{Bichler_2021} model strategies using neural networks~\citep{McCulloch_1943, Rosenblatt_1958}.
Each player's policy network takes as input a player's observation and outputs an action. These policy networks were then trained using \emph{neural pseudogradient ascent}, which uses Gaussian smoothing and applies simultaneous gradient ascent. As the authors note, their policy networks can only model pure strategies, since the output action is deterministic with respect to the input observation.

We also model strategies using neural networks, with one crucial difference: our policy network \(f_\theta\) takes as input the player's observation \(o\) \emph{together with noise} \(z\) from some fixed latent distribution, such as the standard multivariate Gaussian distribution.
Thus the output \(a = f_\theta(o, z)\) of the network is \emph{random} with respect to \(o\). The network can then learn to transform this randomness into a desired action distribution (see Figure~\ref{fig:policynet} in the appendix). This lets us model mixed strategies, which is especially desirable in games that lack pure-strategy equilibria.
Some approaches in the literature use the output of a policy network to parameterize some parametric distribution on the action space, such as a Gaussian mixture model. However, taking the randomness \emph{as input} and letting the neural network \textit{reshape} it as desired allows us to model arbitrary distributions more flexibly.

Figure~\ref{fig:policynet} illustrates the high-level structure of a randomized policy network. It takes as input an observation and random noise, concatenates them, passes the result through a feedforward neural network, and outputs an action.

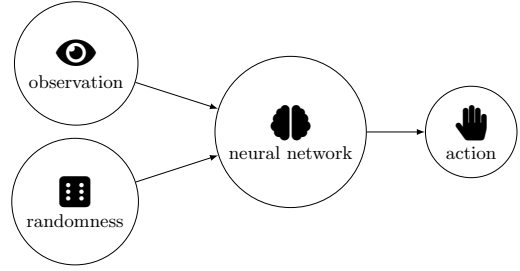
\begin{figure}
\centering
\resizebox{.8\columnwidth}{!}{
\begin{tikzpicture}[circle,align=center]
    \node[draw] (brain) {{\Huge \faBrain}\\neural network};
    \node[draw] (eye) [above left = -.5 and 2 of brain] {{\Huge \faEye}\\observation};
    \node[draw] (dice) [below left = -.5 and 2 of brain] {{\Huge \faDiceSix}\\randomness};
    \node[draw] (hand) [right = of brain] {{\Huge \faHandPaper}\\action};
    \path[->,>=latex]
    (eye) edge (brain)
    (dice) edge (brain)
    (brain) edge (hand)
    ;
\end{tikzpicture}
}
\caption{Structure of a randomized policy network.}
\label{fig:policynet}
\end{figure}

Since the dimensionality of noise fed into a randomized policy network is an important hyperparameter, we now review the literature that studies the relation between input noise dimension and representational power in neural network-based generative models.

The universal approximation theorem~\citep{Cybenko_1989, Hornik_1991, Leshno_1993, Pinkus_1999} states that a single (sufficiently wide) hidden layer suffices to approximate arbitrary continuous functions on a compact domain.
Furthermore, if \(\mathcal{X}, \mathcal{Y}, \subseteq \mathbb{R}^m\), \(X\) is an \(\mathcal{X}\)-valued random variable, \(f : \mathcal{X} \to \mathcal{Y}\), and \(f_n\) is a sequence of functions that converges pointwise to \(f\), the sequence of random variables \(Y_n = f_n(X)\) converges in distribution to \(Y = f(X)\)~\citep[Lemma 4]{Huang_2018}.
Gaussian input noise of \emph{lower} dimension than the output space does not suffice to approximate arbitrary distributions on the output space. In particular, Sard's theorem~\citep{Sard_1942} says the following: 
Let \(f : \mathbb{R}^n \to \mathbb{R}^m\) be \(k\) times continuously differentiable, where \(k \geq \max \{n - m + 1, 1\}\). Let \(X \subseteq \mathbb{R}^n\) be the set of critical points of \(f\) (that is, points where the Jacobian matrix of \(f\) has rank less than \(m\)). Then the image \(f[X]\) has Lebesgue measure zero in \(\mathbb{R}^m\).
As a corollary, if \(n < m\), then all points in \(\mathbb{R}^n\) are critical. Thus \(\operatorname{im} f\) has Lebesgue measure zero in \(\mathbb{R}^m\).\footnote{From a standard uniform random variable, one can extract two independent variables by de-interleaving its binary expansion, but this operation is highly discontinuous and pathological.
}

\citet{Padala_2021} study the effect of input noise dimension in GANs. They show that the right dimension of input noise for optimal results depends on the dataset and architecture used.
\citet{Feng_2021} study how noise injection in GANs helps them overcome the ``adversarial dimension trap'', which arises when
the generated manifold has an intrinsic dimension lower than that of the data manifold: that is, when the latent space is low-dimensional compared to the high-dimensional space of real image details.
Citing Sard's theorem~\citep{Petersen_2006}, they advise against mapping low-dimensional feature spaces into feature manifolds with higher intrinsic dimensions.
\citet{Bailey_2018} investigate the ability of generative networks to convert input noise distributions into other distributions.
One question they study is how easy it is to create a network that outputs \emph{more} dimensions of noise than it receives.
They derive bounds showing that an increase in dimension requires a large, complicated network.
For example, an approximation of the uniform distribution on the unit square using the uniform distribution on the unit interval could use an (almost) space-filling curve such as the iterated tent map. 
(A space-filling curve is a continuous surjection from the unit interval to the unit square or volume of higher dimension.)
This function is highly nonlinear and it can be shown that neural networks must be large to approximate it well.
Thus the dimensionality of input noise is essential in practice. As we discuss later in this paper, our experiments support this conclusion for the game context.
\citet{Bailey_2018} also show that, even when the input dimension is greater than the output dimension, increased input dimension can still sometimes improve accuracy.
For example, to approximate a univariate Gaussian distribution with a high dimensional uniform distribution, one can sum the inputs. By the Berry-Esseen theorem~\citep{Berry_1941}, a refinement of the central limit theorem, the output is close to a Gaussian distribution. This uses no nonlinearity, but simply takes advantage of the fact that projecting a hypercube onto a line results in an approximately Gaussian distribution.
In our experiments, we show an example of excess dimensionality improving performance.

\section{Games used in experiments}
We now describe the games used as benchmarks.
\subsection{Colonel Blotto games}

The Colonel Blotto game is a two-player zero-sum game in which two players distribute resources over several battlefields. A battlefield is won by whoever devotes the most resources to it. A player's payoff is the number of battlefields they win. This game was introduced by~\citet{Borel_1953}. It illustrates fundamental strategic considerations that arise in conflicts or competition involving resource allocation, such as political campaigns, research and development competition (where innovation may involve obtaining a collection of interrelated patents), national security and military and systems defense~\citep{Kovenock_2021}.

\citet{Gross_1950} analyzed a continuous variant in which both players have continuous, possibly unequal budgets. They obtained exact solutions for various special cases, including all 2-battlefield cases and all 3-battlefield cases with equal budgets.
\citet{Washburn_2013} generalized to the case where battlefield values are unequal across battlefields.~\citet{Kovenock_2021} generalized to the case where battlefield values are also unequal across players.~\citet{Adamo_2009} studied a variant in which players have incomplete information about the other player's resource budgets.~\citet{Kovenock_2011} studied a model where the players are subject to incomplete information about the battlefield valuations.
\citet{Adsera_2021} analyzed the natural multiplayer generalization of the continuous Colonel Blotto game.

We describe the case with heterogeneous budgets, heterogeneous battlefield values across both players and battlefields, and several players. Formally, suppose there are \(J\) battlefields. Let \(b_i\) be the budget of player \(i\). Let \(v_{ij}\) be the value to player \(i\) of battlefield \(j\).
Player \(i\)'s action space is the standard \(J\)-simplex dilated by their budget: \(A_i = \{ a_{ij} : \mathbb{R} \mid a_{ij} \geq 0, \textstyle\sum_j a_{ij} = b_i \}\).
Player \(i\)'s reward function is \(r_i(a) = \textstyle\sum_j v_{ij} w_{ij}(a)\) where \(w_{ij}\) is the probability that \(i\) wins \(j\). Ties are broken uniformly at random.

\subsection{Single-item auctions}

An auction is a mechanism by which \emph{items} are sold to \emph{bidders}. 
Auctions play a central role in the study of markets and are used in a wide range of contexts.
In a single-item sealed bid auction, bidders simultaneously submit bids and the highest bidder wins the item. Let \(w_i(a)\) be the probability \(i\) wins given action profile \(a\), where ties are broken uniformly at random.
Let \(v_i(\omega)\) be the item's value for the \(i\)th player given state \(\omega\).
In a \(k\)th-price \emph{winner-pay} auction, the winner pays the \(k\)th highest bid: \(r_i(\omega, a) = w_i(a) (v_i(\omega) - a_{(k)})\), where \(a_{(k)}\) is the \(k\)th highest bid.
In an \emph{all-pay} auction, each player always pays their bid: \(r_i(\omega, a) = w_i(a) v_i(\omega) - a_i\).
This auction is widely used to model lobbying for rents in regulated and trade protected industries, technological competition and R\&D races, political campaigns, job promotions, and other contests~\citep{Baye_1996}.
The all-pay complete-information auction lacks pure-strategy equilibria~\citep{Baye_1996}.
The 2-player 1st-price winner-pay asymmetric-information auction also lacks pure-strategy equilibria~\citep[section 8.3]{Krishna02:Auction}. In particular, the second player must randomize.
More details about each type of auction can be found in the appendix.

\subsection{Multi-item auctions}

Multi-item auctions are of great importance in practice, for example in strategic sourcing~\citep{Sandholm13:Very} and radio spectrum allocation~\citep{Milgrom14:Deferred,Milgrom_2020}.
However, deriving equilibrium bidding strategies for multi-item auctions is notoriously difficult. A rare notable instance where it has been derived is the \emph{chopstick auction}, which was introduced and analyzed by~\citet{Szentes_2003, Szentes03:Chopsticks}. In this auction, 3 chopsticks are sold simultaneously in separate first-price sealed-bid auctions. There are 2 bidders, and it is common knowledge that a pair of chopsticks is worth \$1, a single chopstick is worth nothing by itself, and 3 chopsticks are worth the same as 2. In short, one needs two chopsticks to eat. 
Pure strategies are triples of non-negative real numbers (bids).
This game has an interesting equilibrium: let the tetrahedron \(T\) be defined as the convex hull of the four points \((\tfrac{1}{2}, \tfrac{1}{2}, 0)\), \((\tfrac{1}{2}, 0, \tfrac{1}{2})\), \((0, \tfrac{1}{2}, \tfrac{1}{2})\), and \((0, 0, 0)\).
Then the uniform probability measure on the 2-dimensional surface of \(T\) generates a symmetric equilibrium.
(Furthermore, all points inside the tetrahedron are pure best responses to this equilibrium mixture.) 
We benchmark on the chopsticks auction since it is a rare case of a multi-item auction where the solution can be checked because the equilibrium is known. It is also a canonical case of simultaneous separate auctions under combinatorial preferences.

\subsection{Visibility game}

\citet{Lotker_2008} introduced the \emph{visibility game}, a noncooperative, complete-information strategic game. In this game, each player \(i\) chooses a point \(x_i \in [0, 1]\). Their payoff is the distance to the next higher point, or to \(1\) if \(x_i\) is the highest.
This game models a situation where players seek to maximize their \emph{visibility time}, and is a variant of the family of ``timing games''~\citep{Fudenberg91:Gamea}.
It resembles the ``war of attrition'' game formalized by~\citet{Smith74:theory}.
In this game, both players are engaged in a costly competition and they need to choose a time to concede. More formally, the first player to concede (called ``leader'') gets a smaller payoff than the other player (called ``follower''). Furthermore, the payoff to the leader strictly decreases as time progresses. That is, conceding early is better than conceding late.

They prove that the \(n\)-player visibility game has no pure equilibrium, but has a unique mixed equilibrium, which is symmetric. In the 2-player case, up to a set of measure zero, there is a unique equilibrium whose strategies have probability densities \(p(x) = 1/(1-x)\) when \(0 \leq x \leq 1 - 1/\mathrm{e}\) and 0 otherwise. Each player's expected payoff is \(1/\mathrm{e}\).

\section{Hyperparameters used in the experiments}

To initialize our networks, we use He initialization~\citep{He_2015}, which is widely used for feedforward networks with ReLU-like activation functions. It initializes bias vectors to zero and weight matrices with normally-distributed entries scaled by \(\sqrt{2 / n}\), where \(n\) is the layer's input dimension. We use the ELU activation function~\citep{Clevert_2016} for hidden layers. Like~\citet{Bichler_2021}, we use 2 hidden layers with 10 neurons each.

We illustrate the performance of our method by plotting NashConv against the number of epochs, where each epoch consists of \(10^6\) optimization steps. Each hyperparameter setting is labeled in the legend and shown in a different color. Each individual setting is run 20 times with different random seeds. Solid lines indicate means across trials. Bands indicate a confidence interval for this mean with a confidence level of 0.95. These confidence intervals are computed using bootstrapping~\citep{Efron_1979}, specifically the bias-corrected and accelerated (BCa) method~\citep{Efron_1987}.

For the gradient estimator, we use the Gaussian distribution with scale \(\sigma = 10^{-2}\), \(N = 1\) samples, and the central-difference stencil (so 2 evaluations per step). For the optimizer, we use standard gradient descent with a learning rate of \(10^{-6}\).
To estimate NashConv (Equation~\ref{eq:br}), we use 100 observation samples and 300 state samples (given each observation). We use a 100-point discretization of the action space for the auctions and visibility game. 
For Colonel Blotto games, we use a 231-point discretization of the action space. It is obtained by enumerating all partitions of the integer 20 into 3 parts and renormalizing them to sum to 1.

\section{Experimental results}

We now describe our experimental results for each game. Figures illustrating analytically-derived equilibria in cases where they are known can be found in the appendix.

\subsection{Colonel Blotto games}

\begin{figure}
    \centering
    \includegraphics{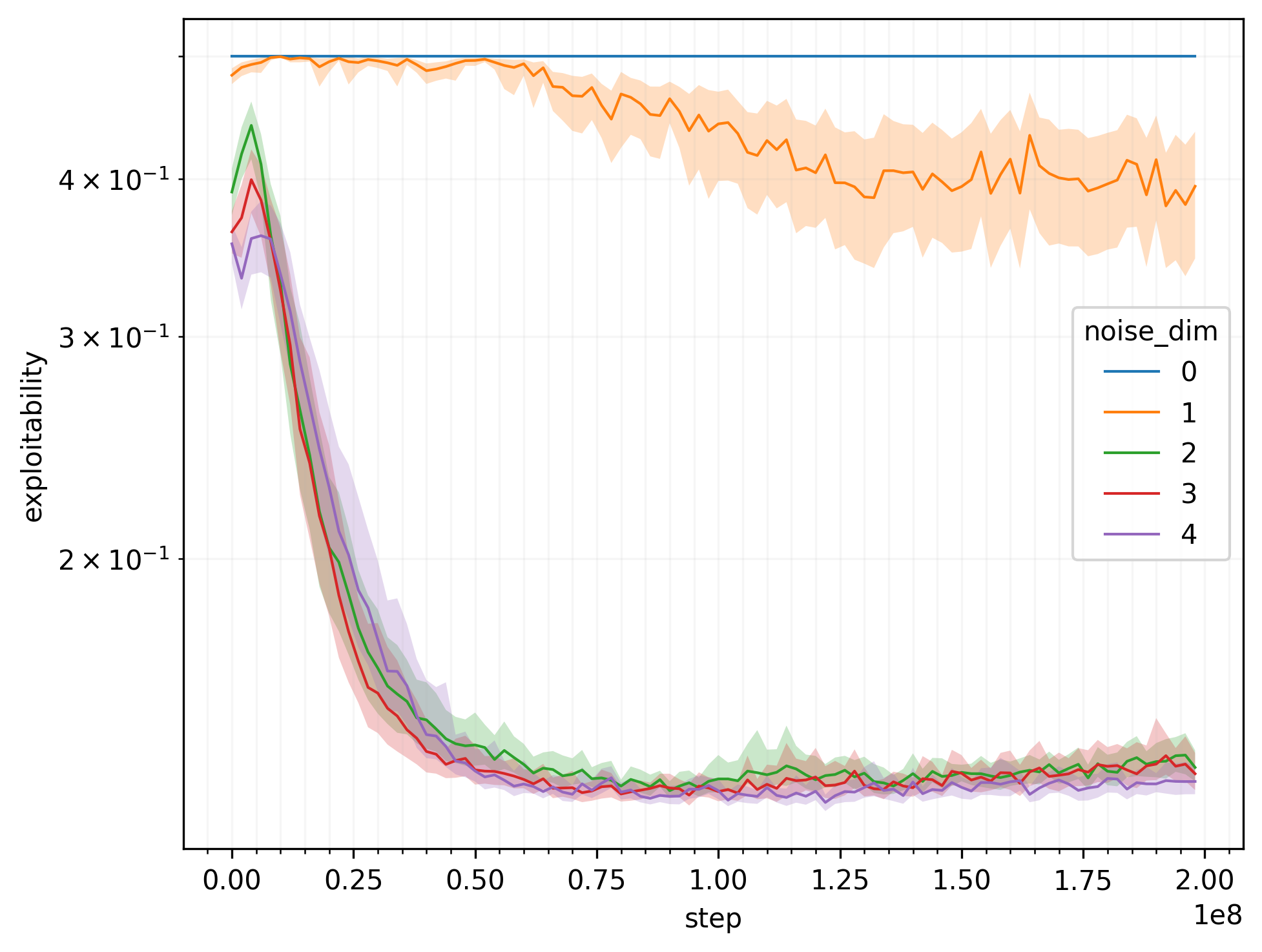}
    \caption{Continuous Colonel Blotto game.}
    \label{fig:blotto_nashconv}
\end{figure}

Actions in the continuous Colonel Blotto game are points on the standard simplex. Thus we use a softmax activation function for the output layer of the randomized policy network. Figure~\ref{fig:blotto_nashconv} illustrates the performance of our method on the continuous Colonel Blotto game with 2 players and 3 battlefields. Since the game has no pure-strategy Nash equilibrium, deterministic strategies perform badly, as expected. 1-dimensional noise results in slightly better performance, but does not let players randomize well enough to approximate the equilibrium. On the other hand, noise of dimension 2 and higher is sufficient for good performance.
The very slight increase in exploitability after 1e8 steps is most likely due to fluctuations introduced by the many sources of stochasticity in the training process, including the game and gradient estimates, as well as the fact that we are training a multi-layer neural network. Even in the supervised learning setting, loss does not always decrease monotonically.

\begin{figure}
    \centering
    \includegraphics[width=.25\columnwidth]{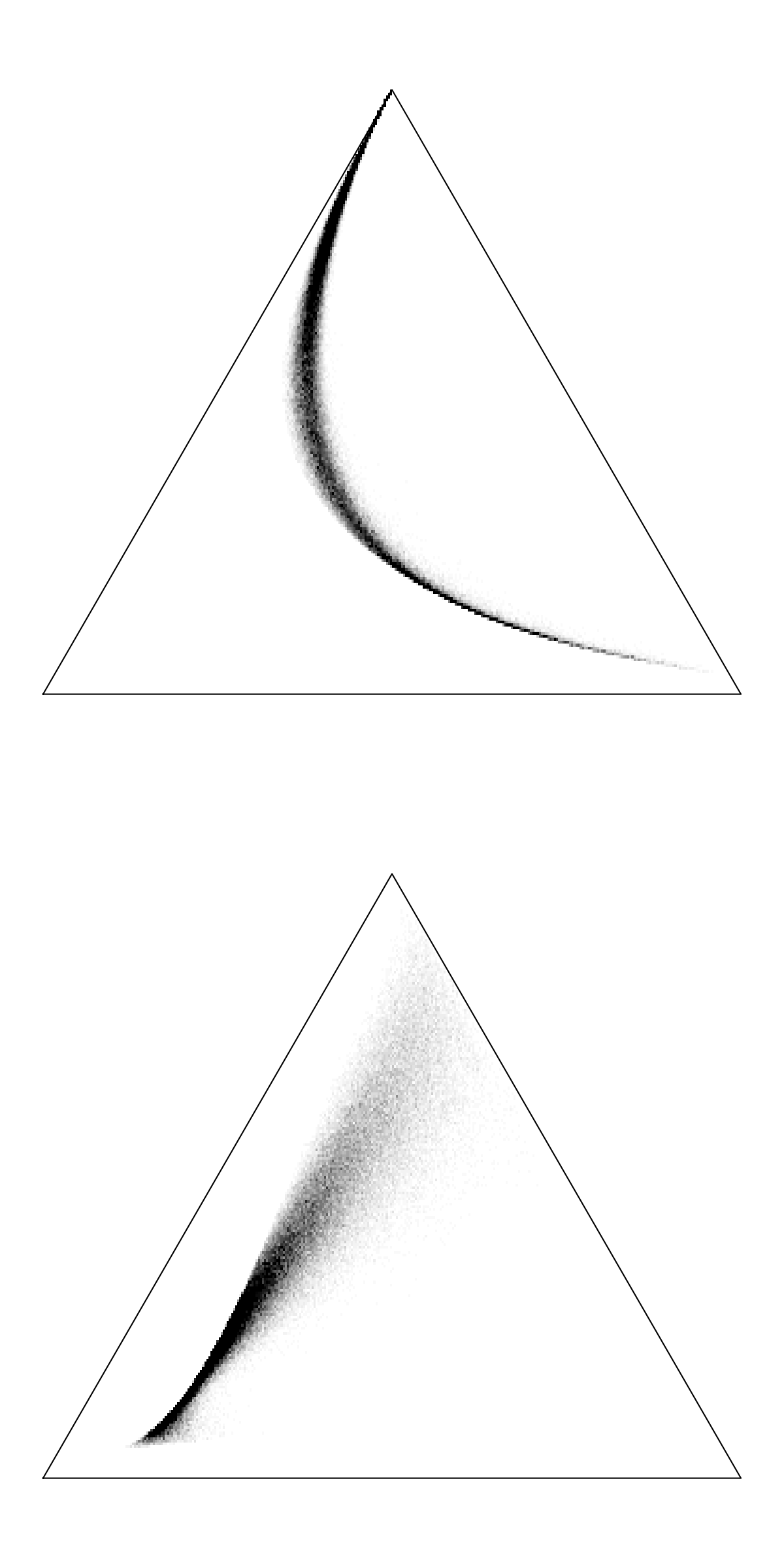}\hfil
    \includegraphics[width=.25\columnwidth]{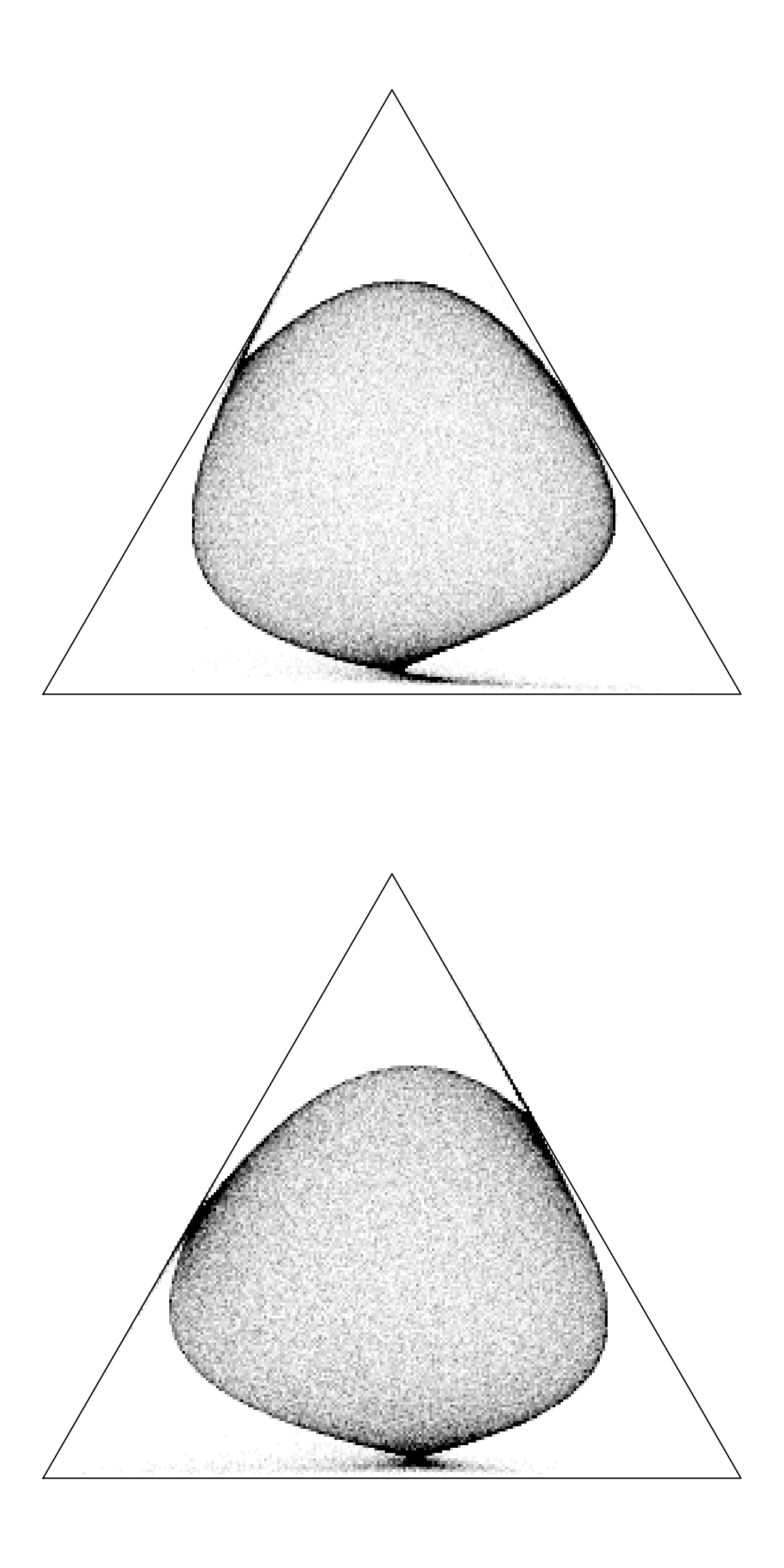}\hfil
    \includegraphics[width=.25\columnwidth]{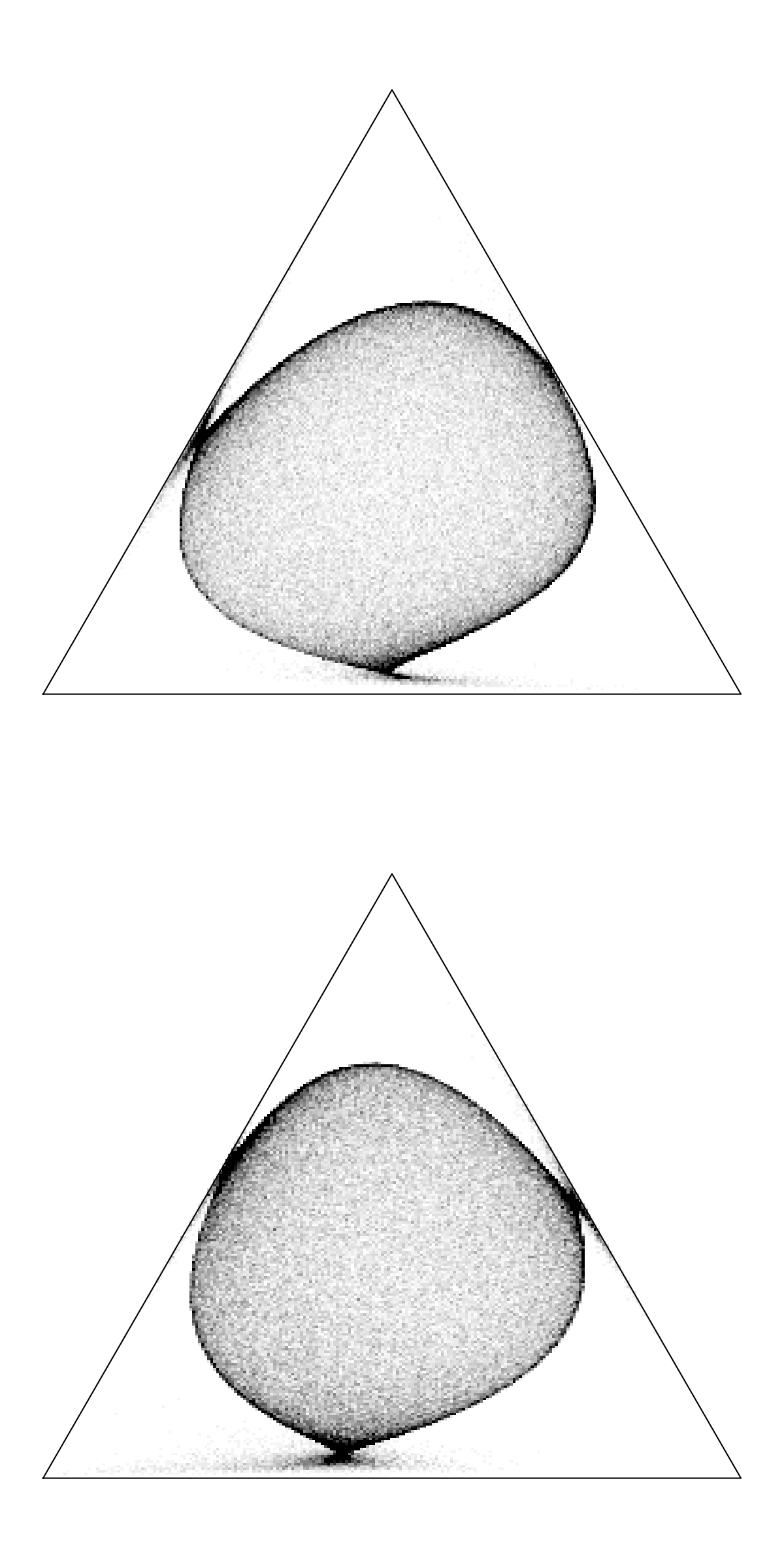}\hfil
    \includegraphics[width=.25\columnwidth]{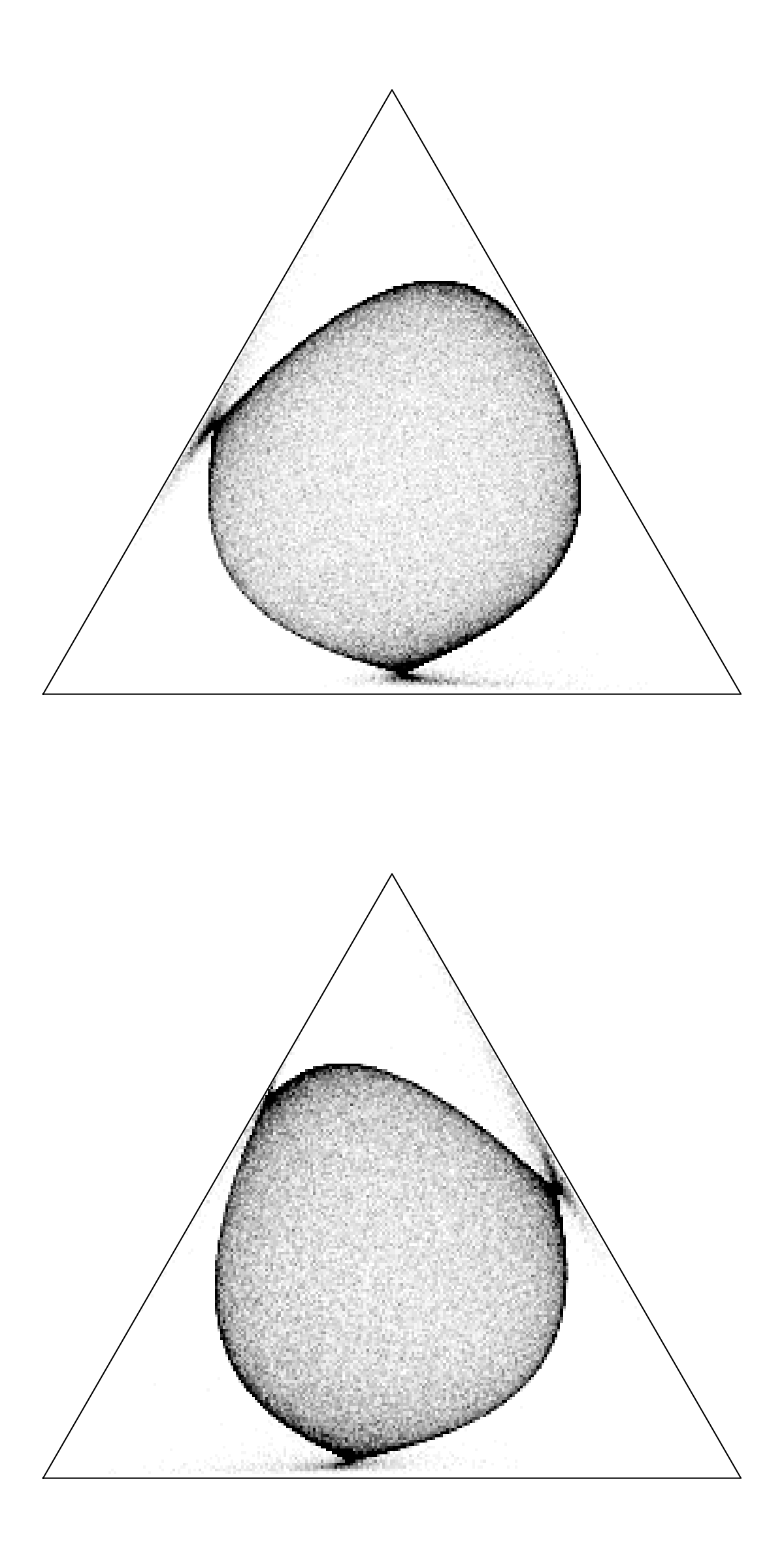}
    \caption{Strategies at epochs 0, 60, 120, and 180 (left to right). Each histogram uses \(10^4\) action samples.}
    \label{fig:blotto_strategies}
\end{figure}

Figure~\ref{fig:blotto_strategies} illustrates the strategies at different stages of training for one trial that uses 2-dimensional noise. Each scatter plot is made by sampling \(10^5\) actions from each player's strategy. The strategies converge to the analytical solution derived by~\citet{Gross_1950}. More details about this solution can be found in the appendix.

\begin{figure}
    \centering
    \includegraphics{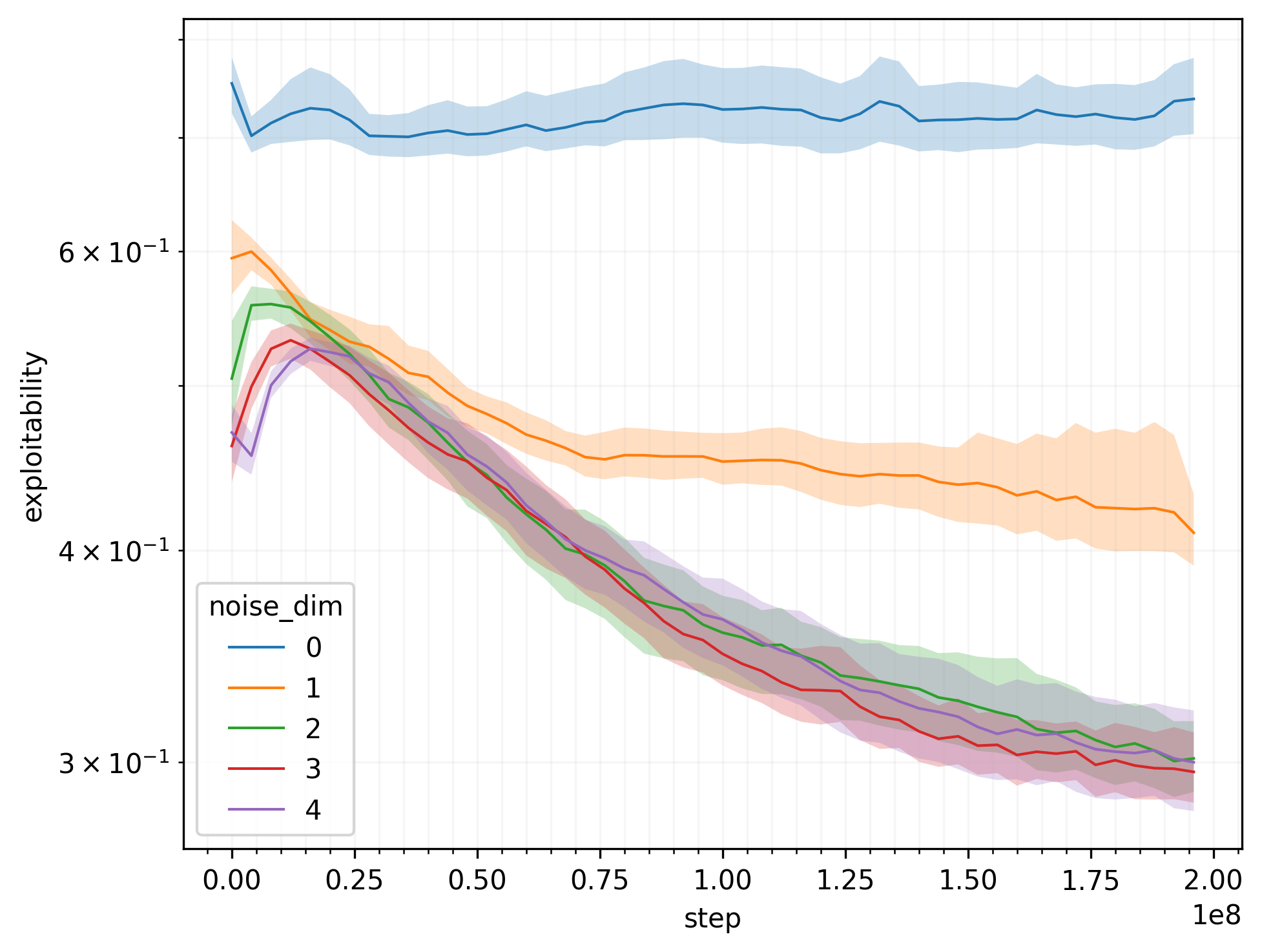}
    \caption{Continuous Colonel Blotto game with random budgets.}
    \label{fig:blotto_nashconv_2}
\end{figure}

Figure~\ref{fig:blotto_nashconv_2} also illustrates performances on the continuous Colonel Blotto game with 2 players and 3 battlefields. This time, however, the budgets for each player are sampled from the standard uniform distribution and revealed to both players. Thus each player must adjust their action distribution accordingly. To our knowledge, prior approaches~\citep{Adam_2021, Kroupa_2021, Ganzfried_2021} do not learn strategies that can generalize across different parameters (like budgets and valuations), which requires the use of function approximators such as neural networks.

\subsection{Single-item and multi-item auctions}

We now turn to the auction setting.
Unlike~\citet{Bichler_2021}, we use an absolute value function in the output layer rather than a ReLU function. The reason is that, as we found in our experiments, ReLU can easily cause degenerate initializations: if the randomly-initialized neural network happens to map all of the unit interval (the observation space) to negative bids, no gradient signal can be received and the network is stuck. By default, auctions are 2-player, 1st-price, and winner-pay unless otherwise noted.

\begin{figure}
    \centering
    \includegraphics{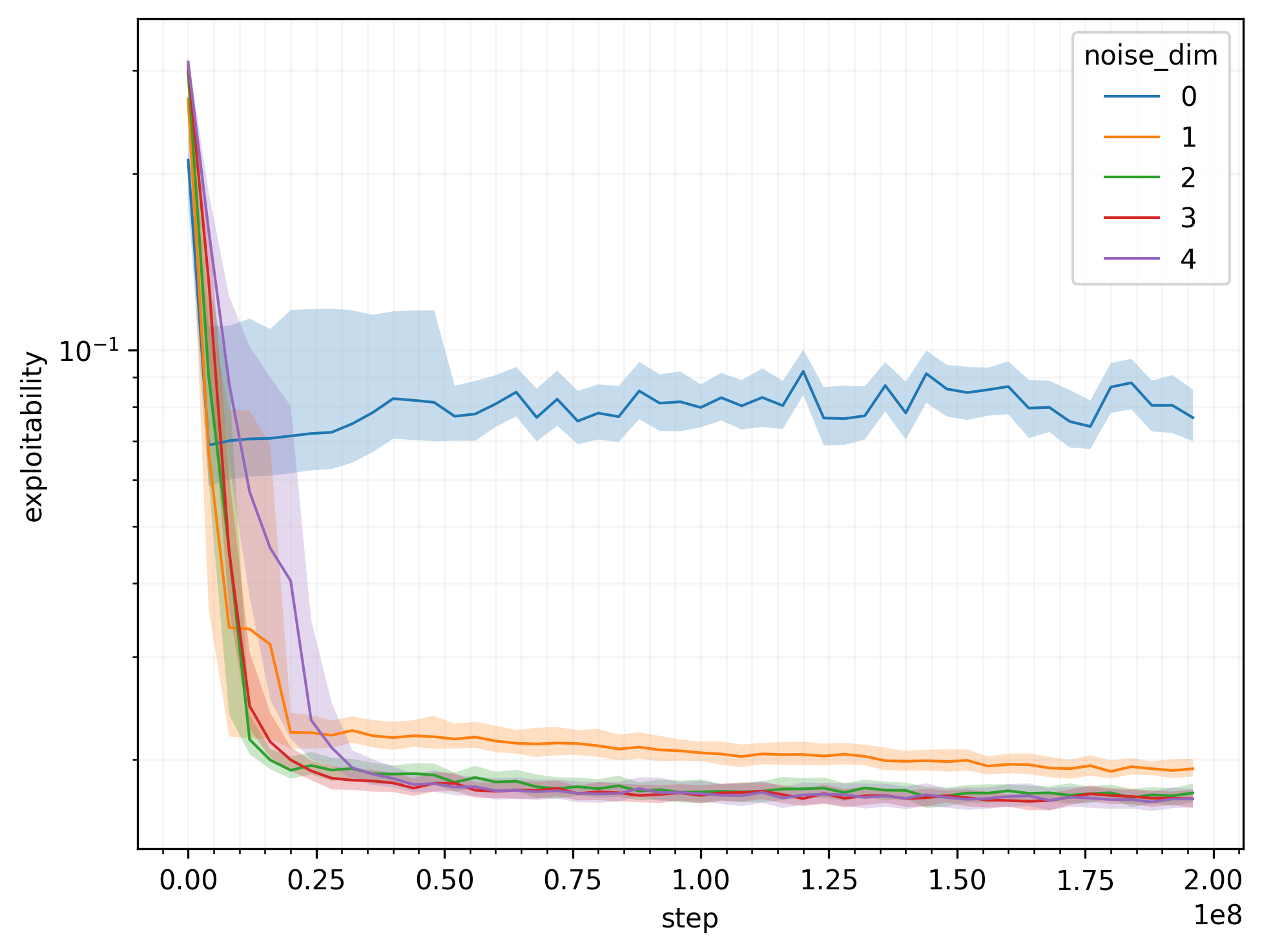}
    \caption{The asymmetric-information auction.}
    \label{fig:asymmetric_nashconv}
\end{figure}

\begin{figure}
    \centering
    \includegraphics[width=.25\columnwidth]{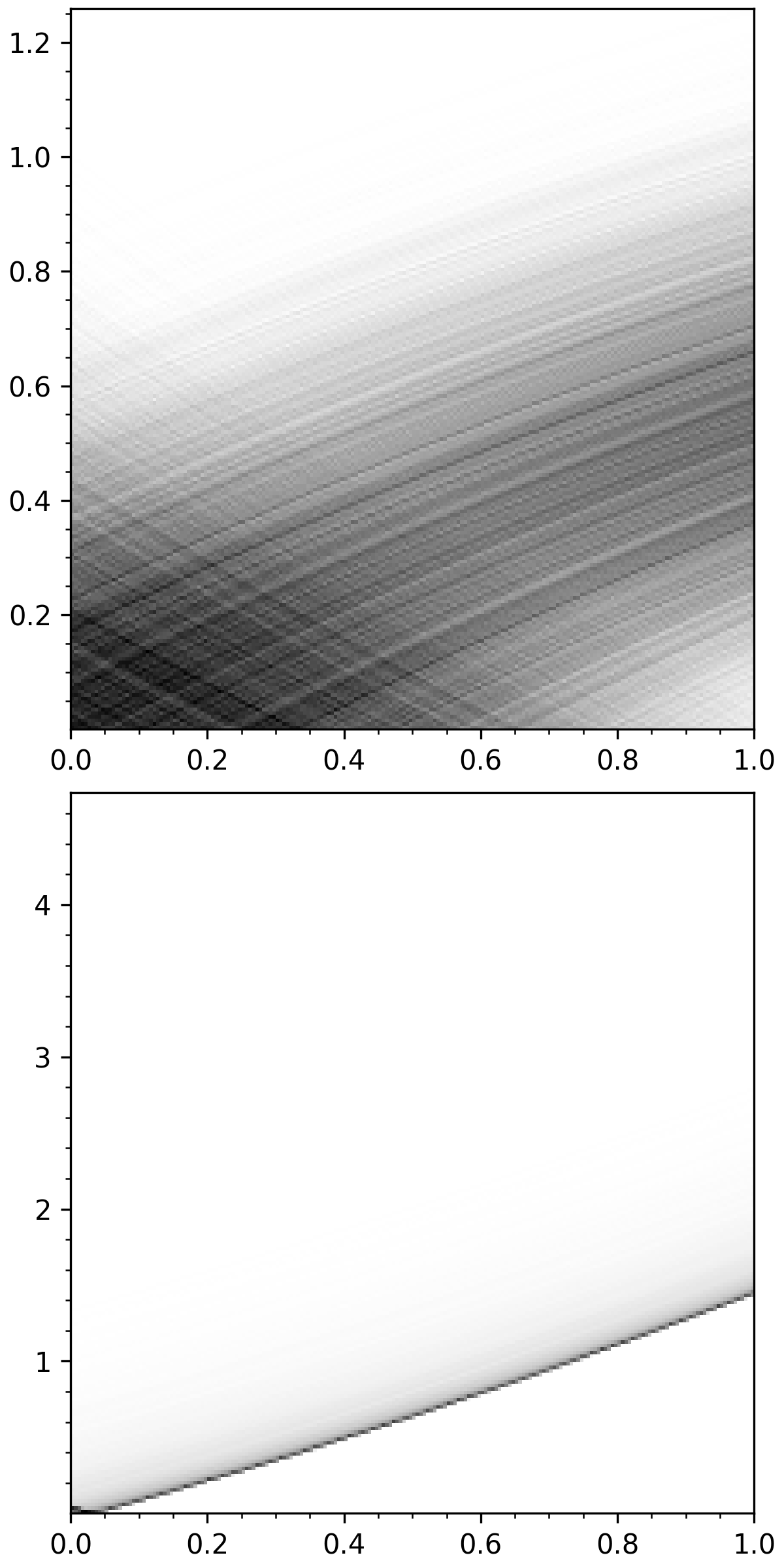}\hfil
    \includegraphics[width=.25\columnwidth]{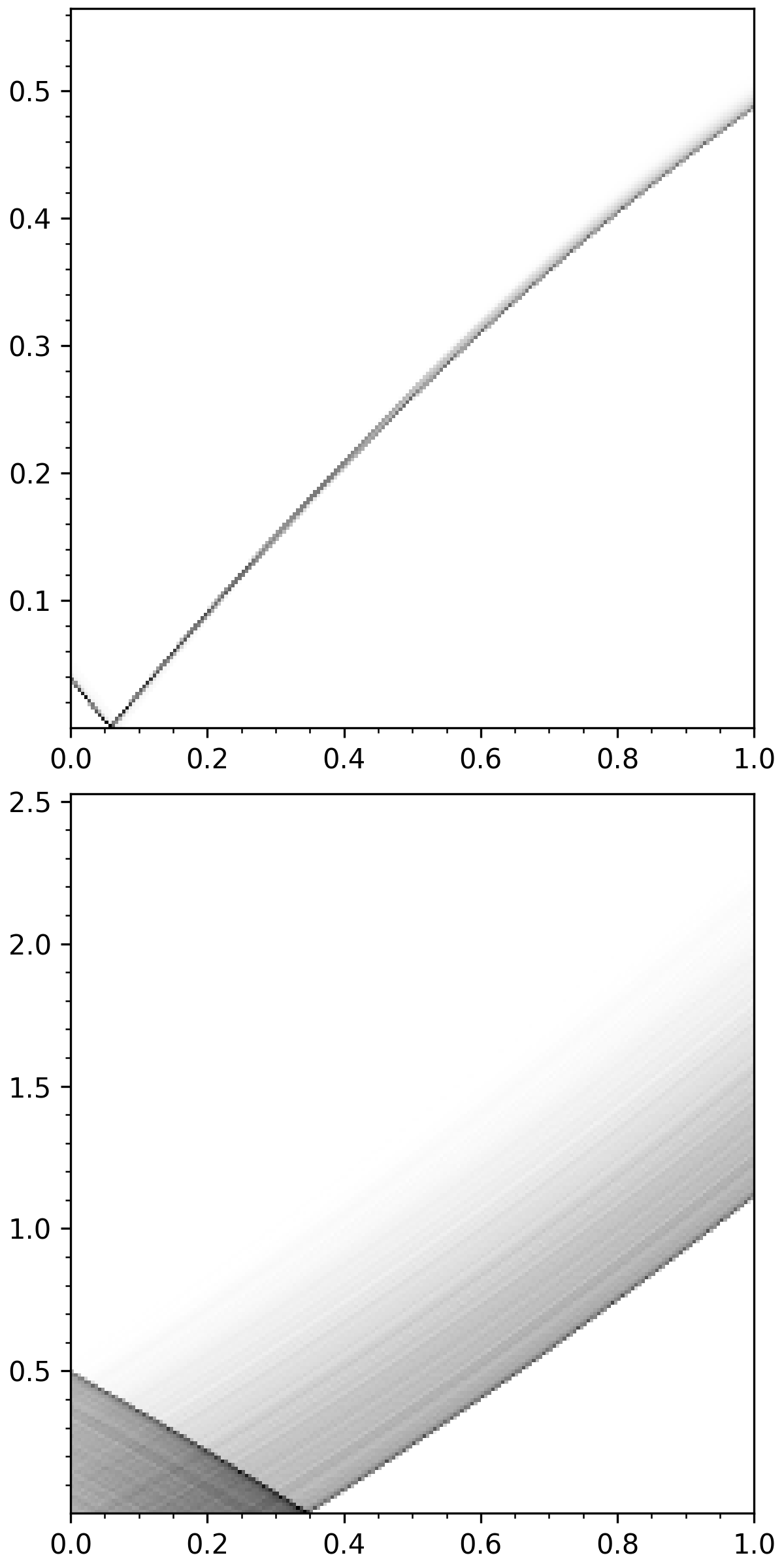}\hfil
    \includegraphics[width=.25\columnwidth]{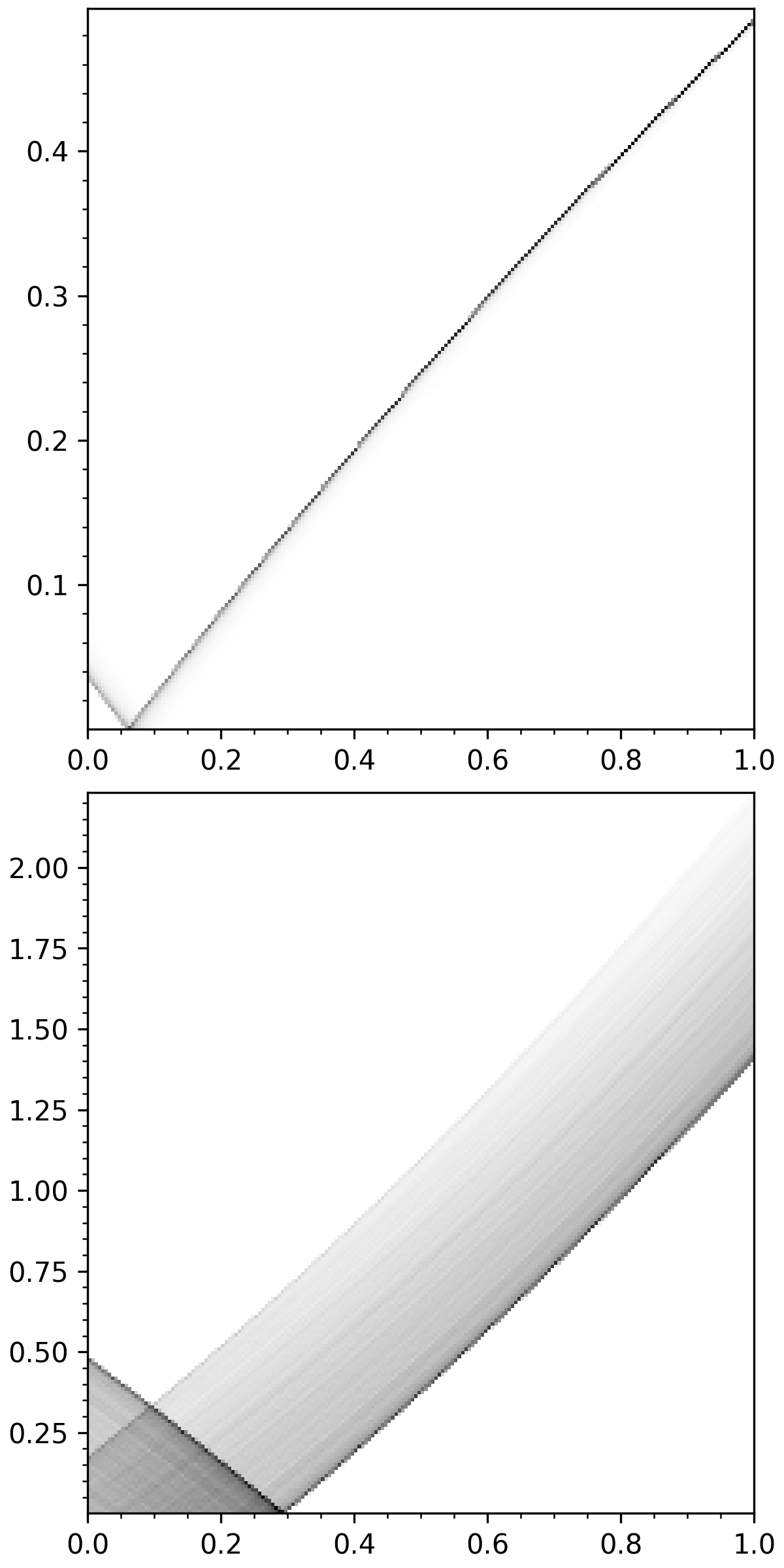}\hfil
    \includegraphics[width=.25\columnwidth]{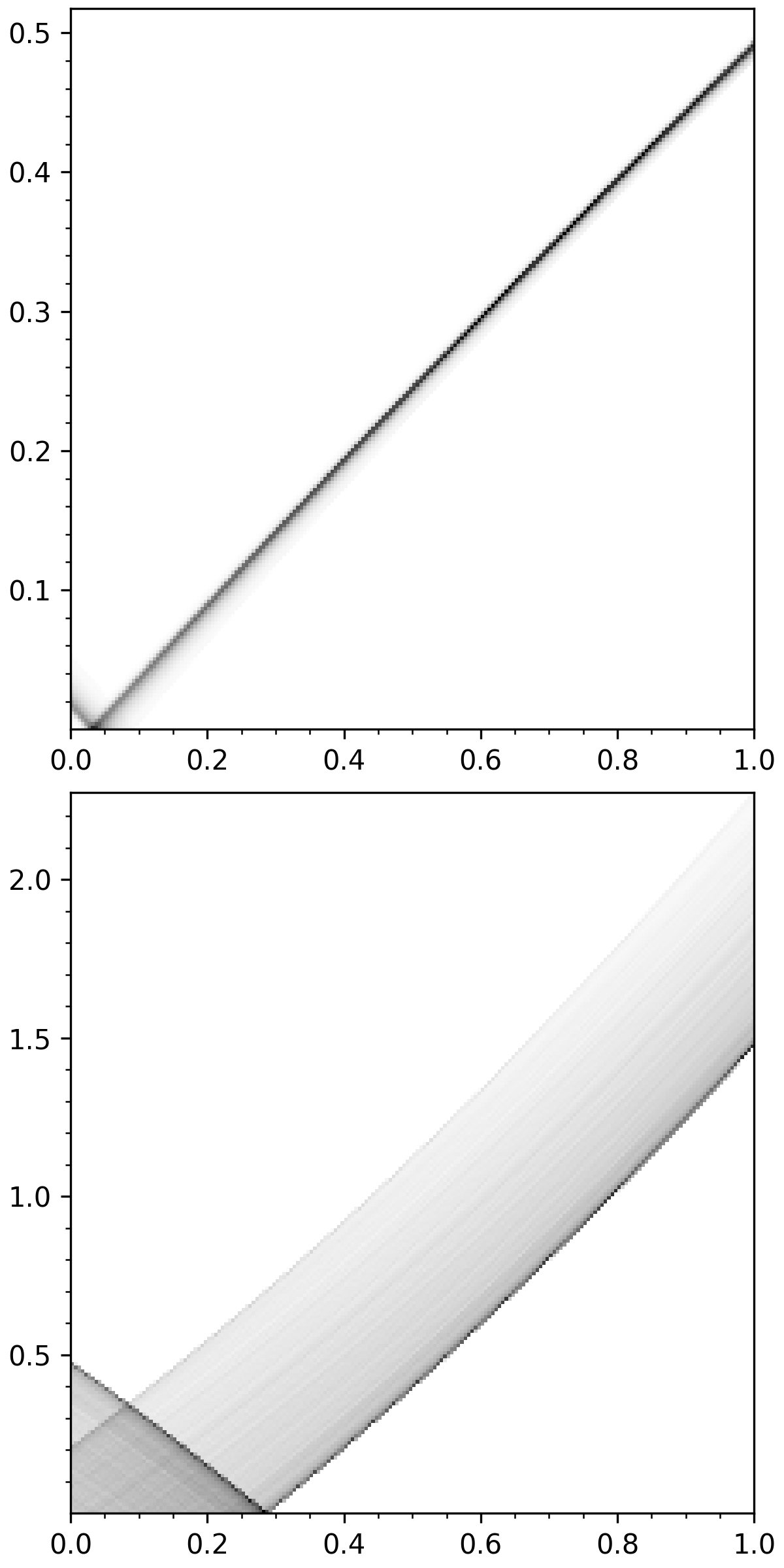}
    \caption{Strategies at epochs 0, 30, 60, and 90 (left to right). X and Y axes denote observation and bid, respectively. Each histogram uses \(10^4\) action samples per observation.}
    \label{fig:asymmetric_strategies}
\end{figure}

\begin{figure}
    \centering
    \includegraphics{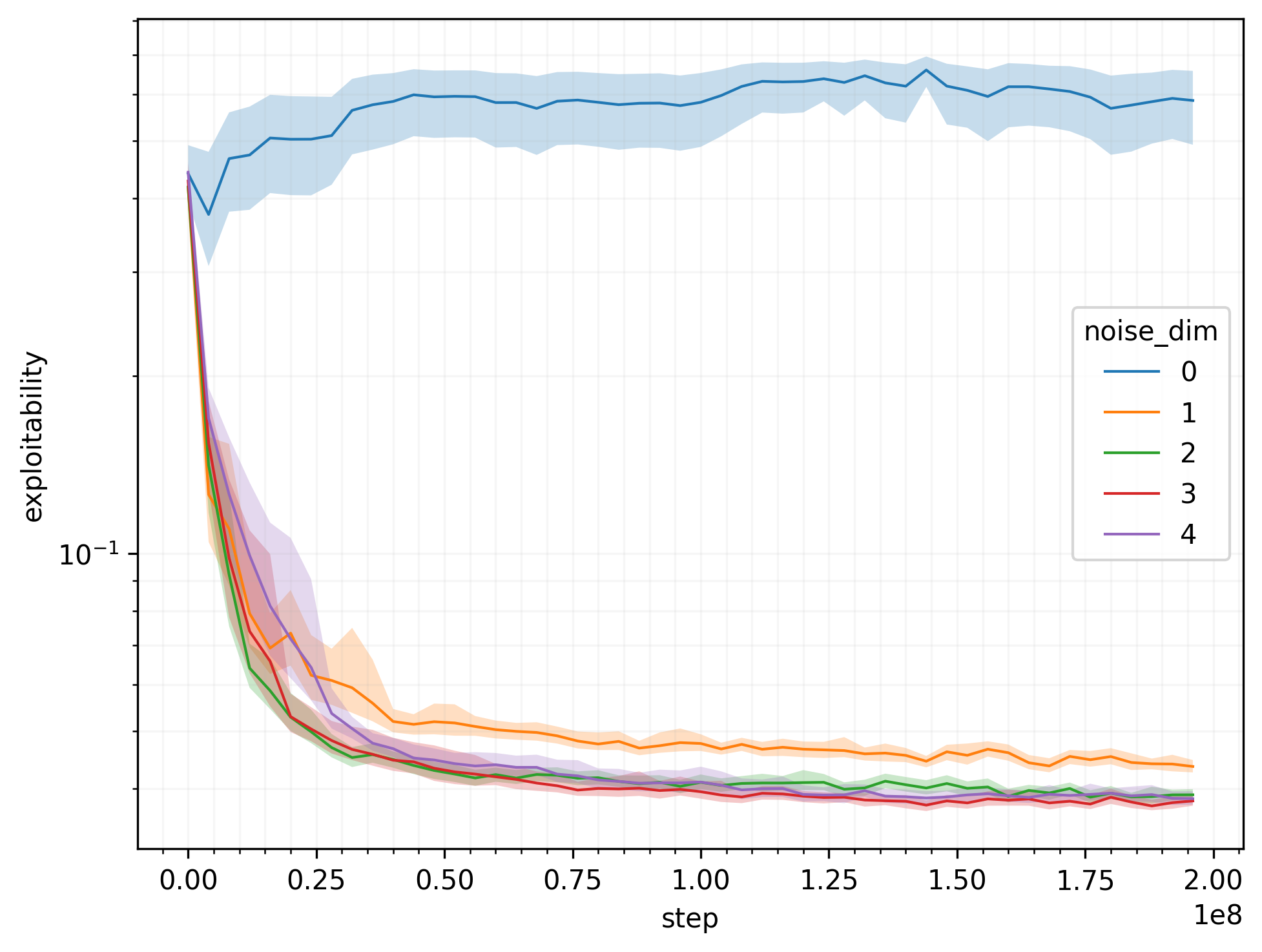}
    \caption{The all-pay complete-information auction.}
    \label{fig:complete_nashconv}
\end{figure}

\begin{figure}
    \centering
    \includegraphics[width=.25\columnwidth]{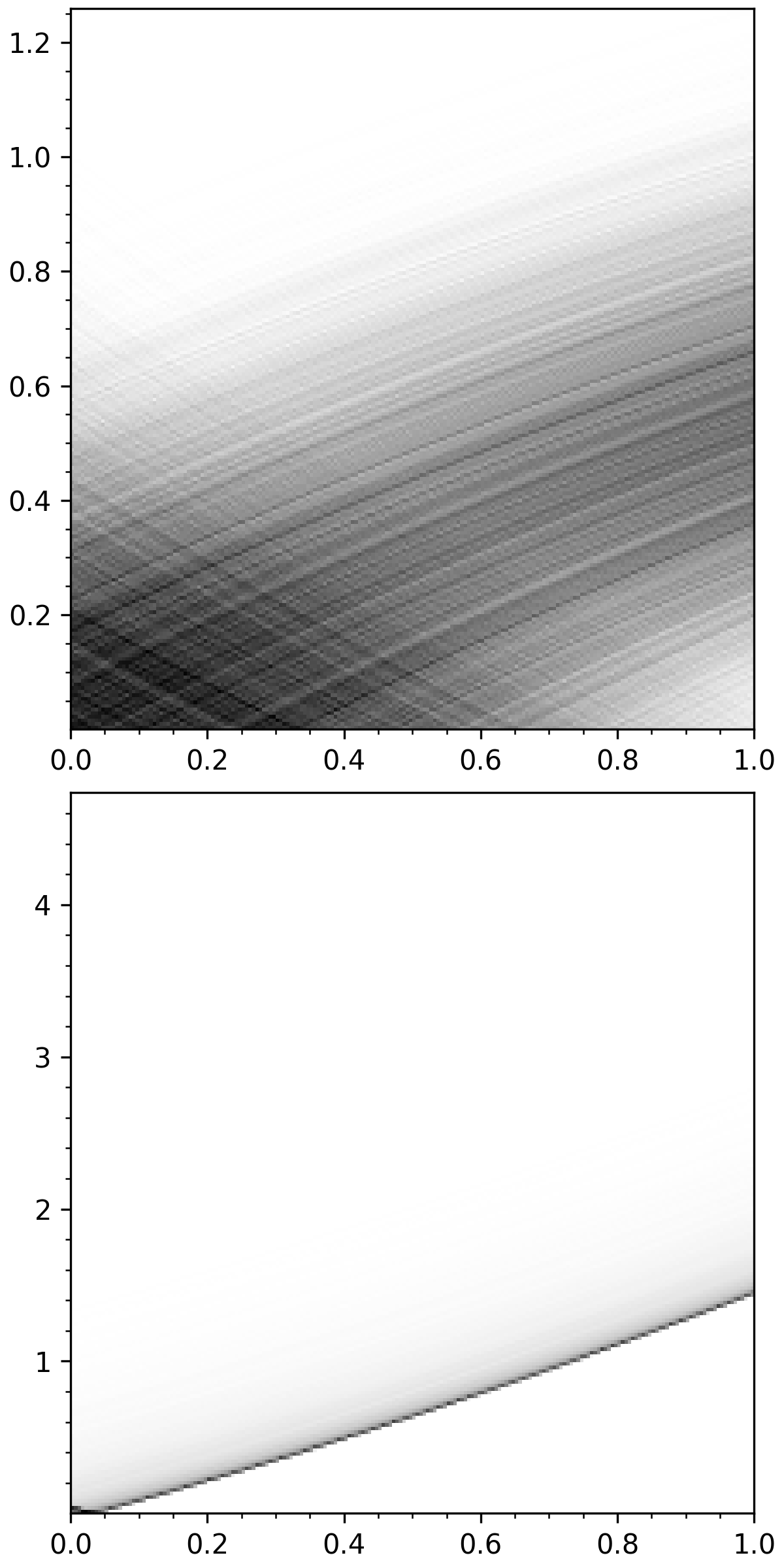}\hfil
    \includegraphics[width=.25\columnwidth]{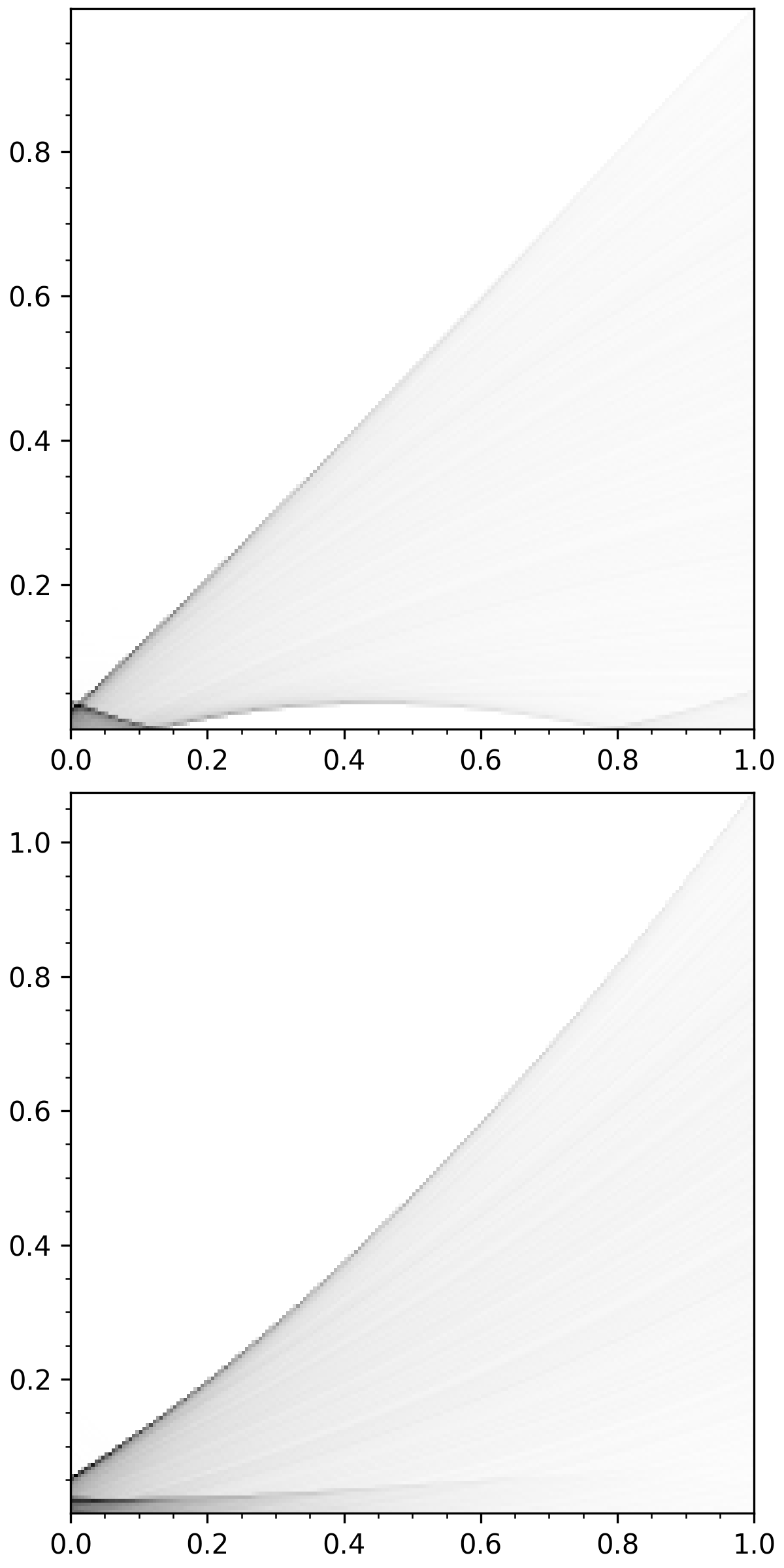}\hfil
    \includegraphics[width=.25\columnwidth]{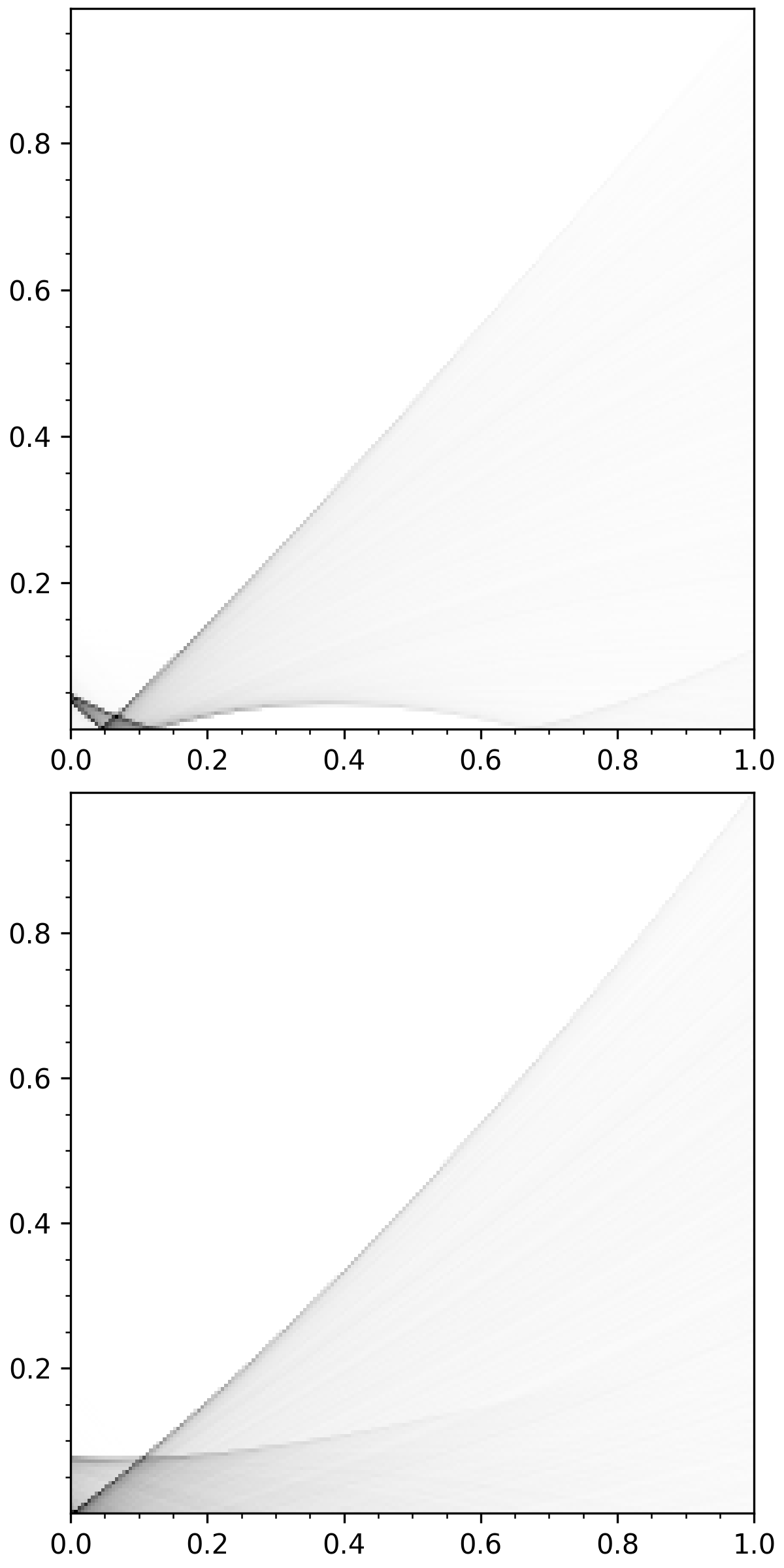}\hfil
    \includegraphics[width=.25\columnwidth]{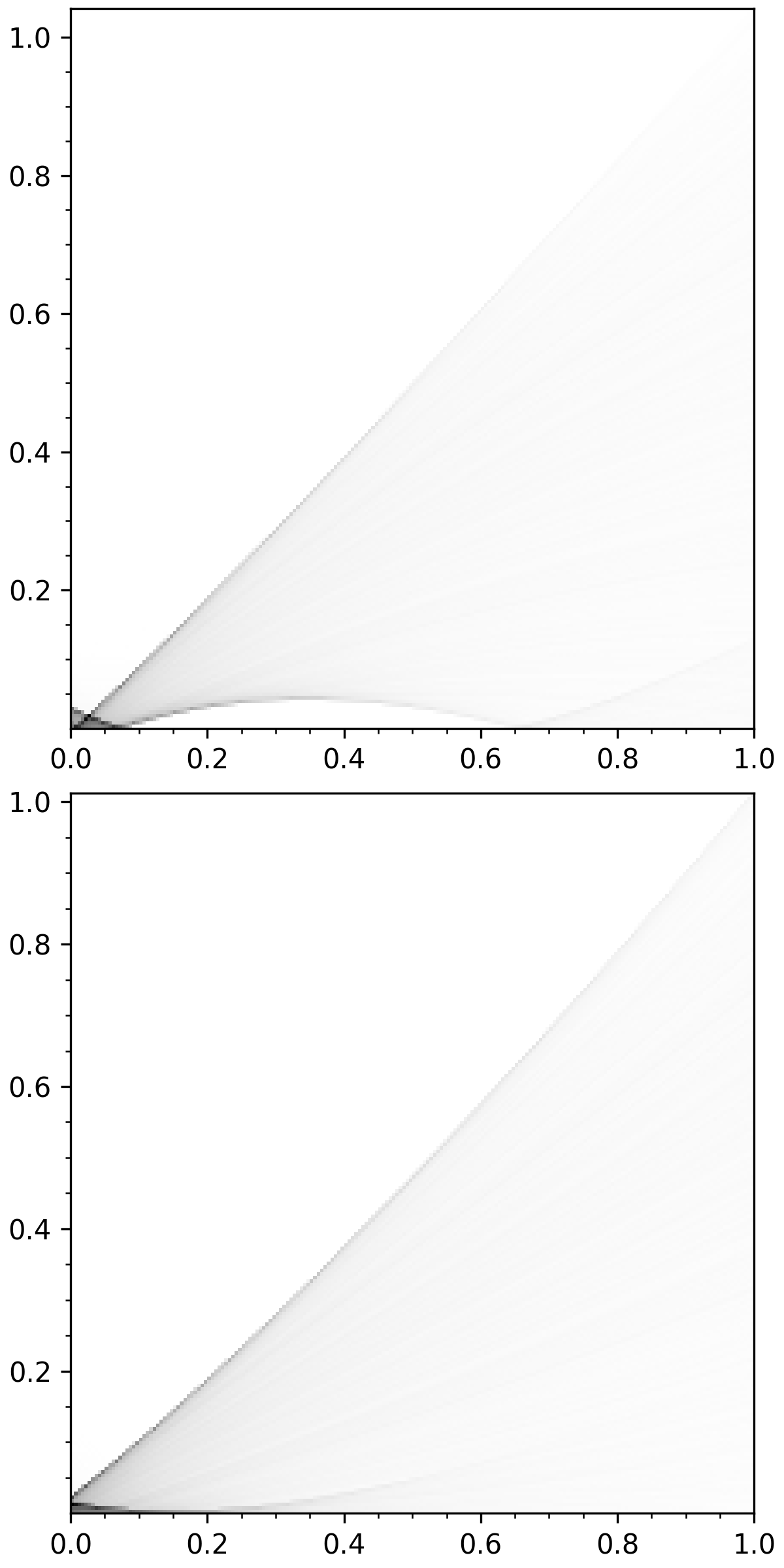}
    \caption{Strategies at epochs 0, 30, 60, and 90 (left to right). X and Y axes denote observation and bid, respectively. Each histogram uses \(10^4\) action samples per observation.}
    \label{fig:complete_strategies}
\end{figure}

Figures~\ref{fig:asymmetric_nashconv} and~\ref{fig:asymmetric_strategies} illustrate performances and strategies for the asymmetric information auction.
Figures~\ref{fig:complete_nashconv} and~\ref{fig:complete_strategies} illustrate performances and strategies for the complete-information all-pay auction.
Recall that these auctions have no pure-strategy equilibria. Thus, as expected, deterministic strategies perform poorly.
As with Colonel Blotto games, our experiments in these auction settings show that the ability to flexibly model mixed strategies is crucial for computing approximate Nash equilibria in certain auction settings.

\begin{figure}
    \centering
    \includegraphics{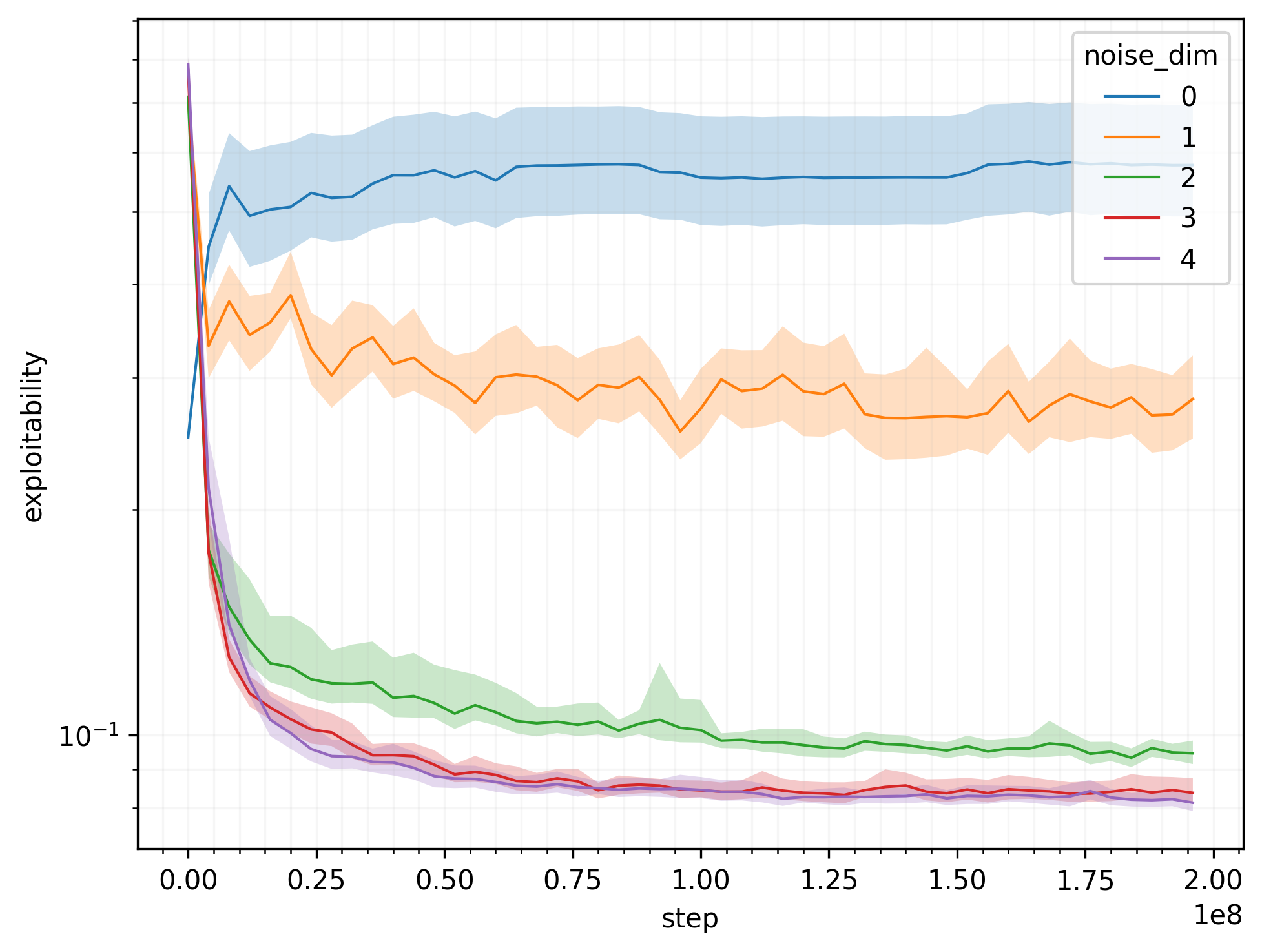}
    \caption{The chopstick auction.}
    \label{fig:chopstick_nashconv}
\end{figure}

\begin{figure}
    \centering
    \includegraphics{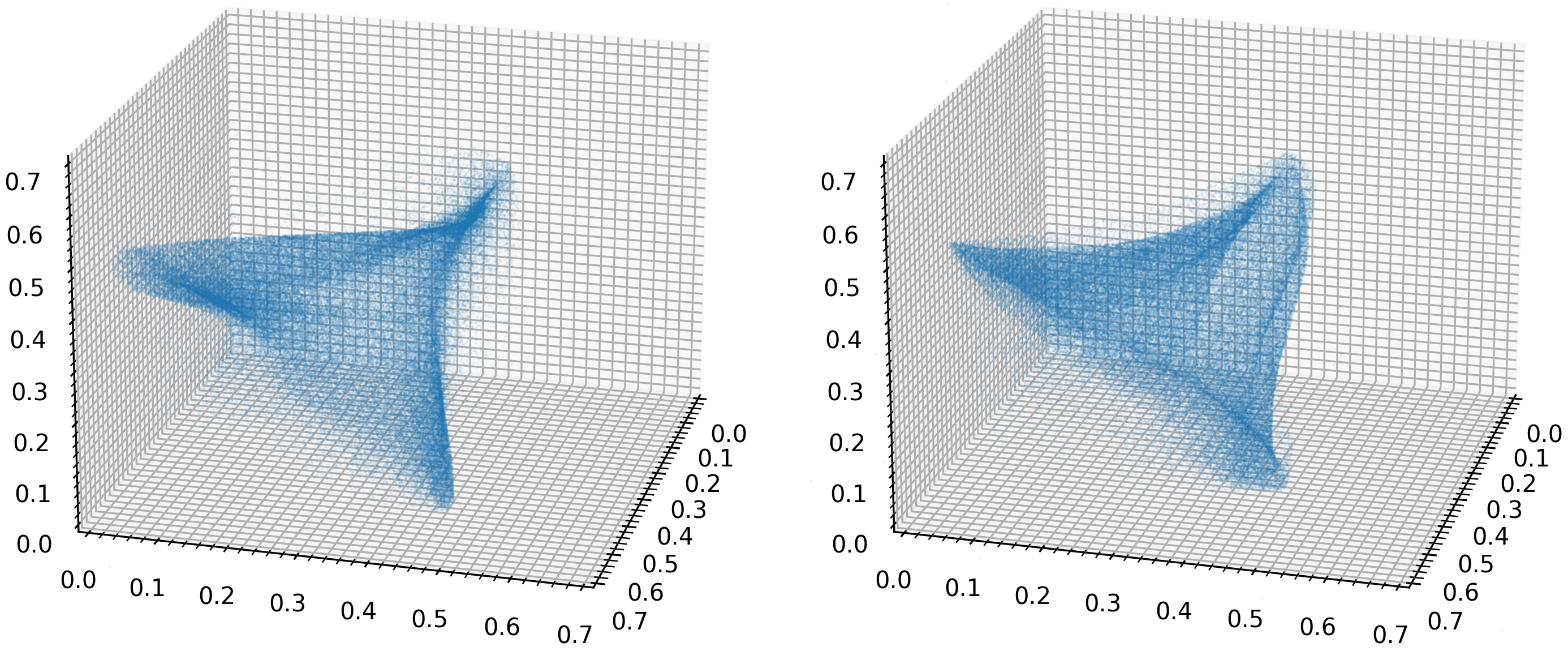}
    \includegraphics{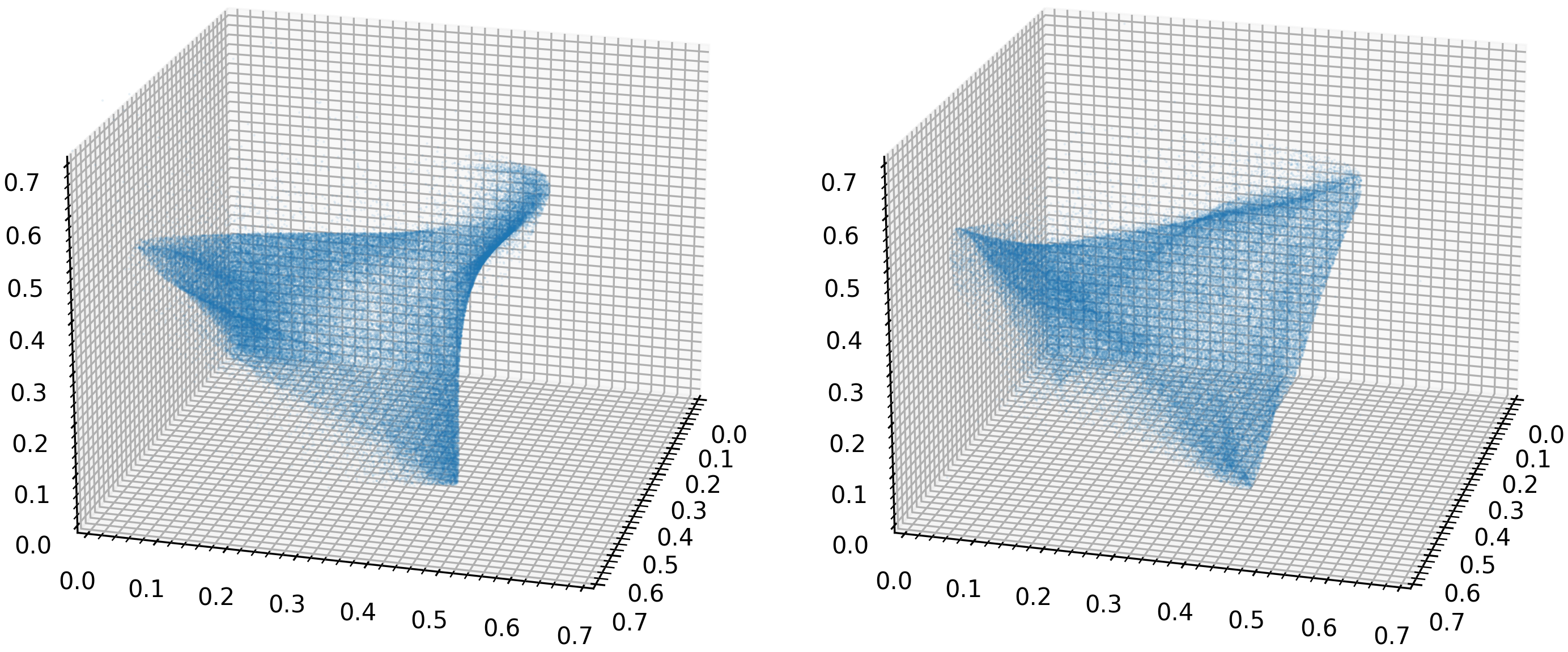}
    \includegraphics{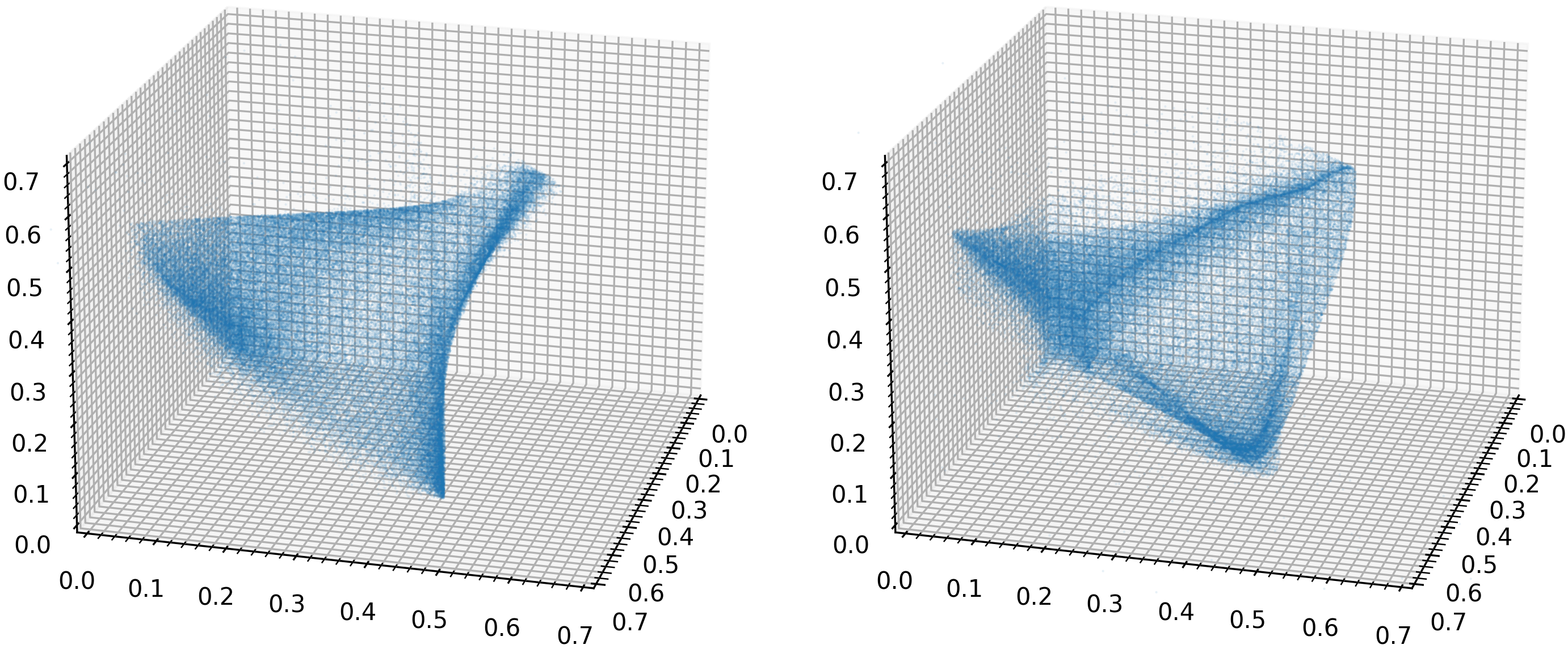}
    \caption{Illustration of player strategies, based on \(10^5\) action samples. Left to right: Players 1 and 2. Top to bottom: Epochs 30, 60, and 90.}
    \label{fig:chopstick_strategies}
\end{figure}

Figure~\ref{fig:chopstick_nashconv} and~\ref{fig:chopstick_strategies} illustrate performances and strategies for the chopstick auction.
Here we encounter an interesting phenomenon: Recall that this game has a symmetric equilibrium generated by the uniform measure on the surface of a tetrahedron. Although the tetrahedron itself is 3-dimensional, its surface is only 2-dimensional. Thus one may wonder whether 2-dimensional noise is sufficient: that is, whether the network can learn to project this lower-dimensional manifold out into the third dimension while ``folding'' it in the way required to obtain the surface of the tetrahedron. Through our experiments, we observe that 2-dimensional noise indeed suffices to (approximately) match the performance of higher-dimensional noise. Thus the \emph{intrinsic dimension} of the equilibrium action distribution (as opposed to the extrinsic dimension of the ambient space it is embedded in) seems to be the decisive factor.

\subsection{Visibility game}

\begin{figure}
    \centering
    \includegraphics{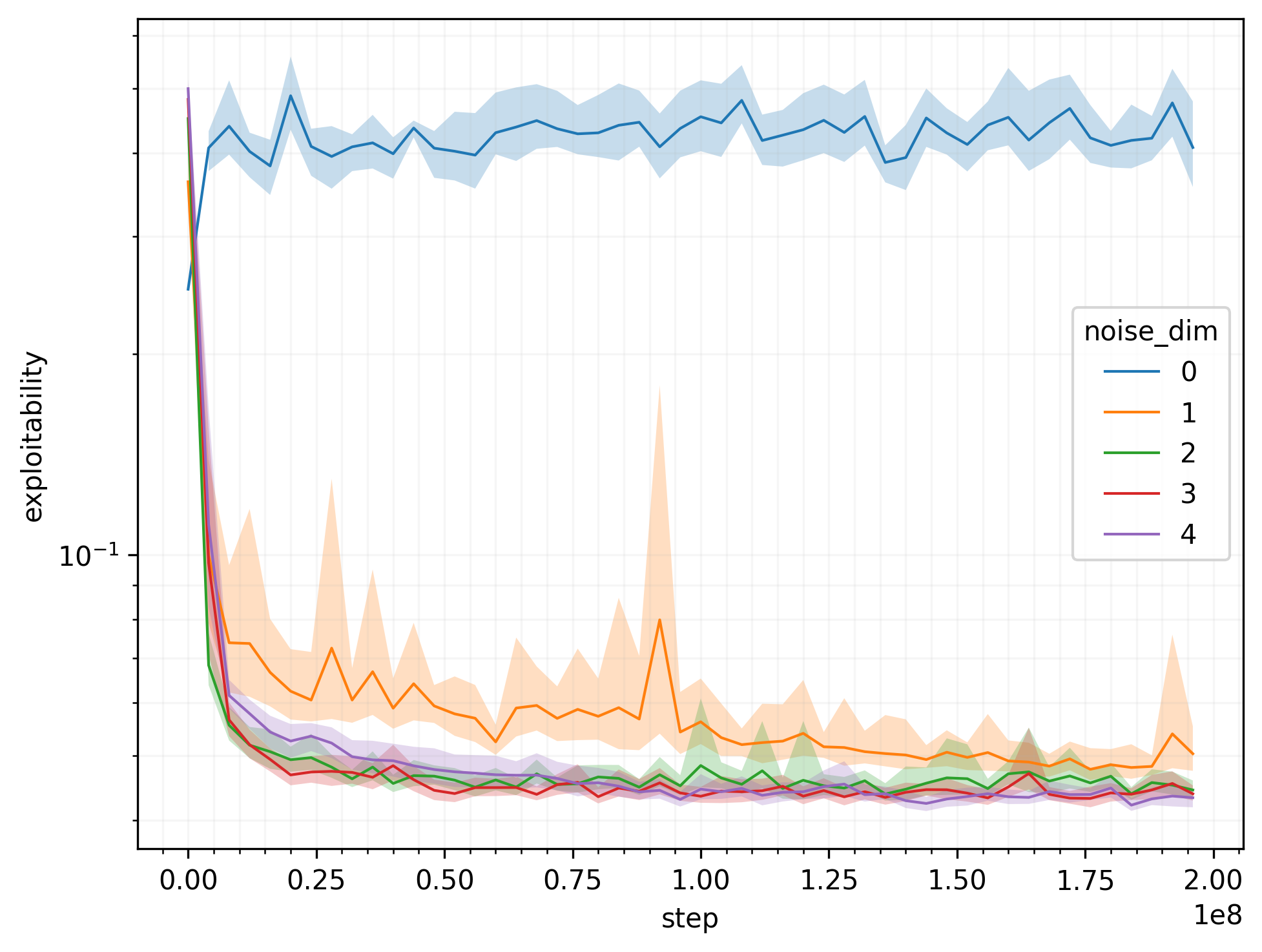}
    \caption{Performance on the 2-player visibility game.}
    \label{fig:visibility_nashconv}
\end{figure}

\begin{figure}
    \centering
    \includegraphics[width=.25\columnwidth]{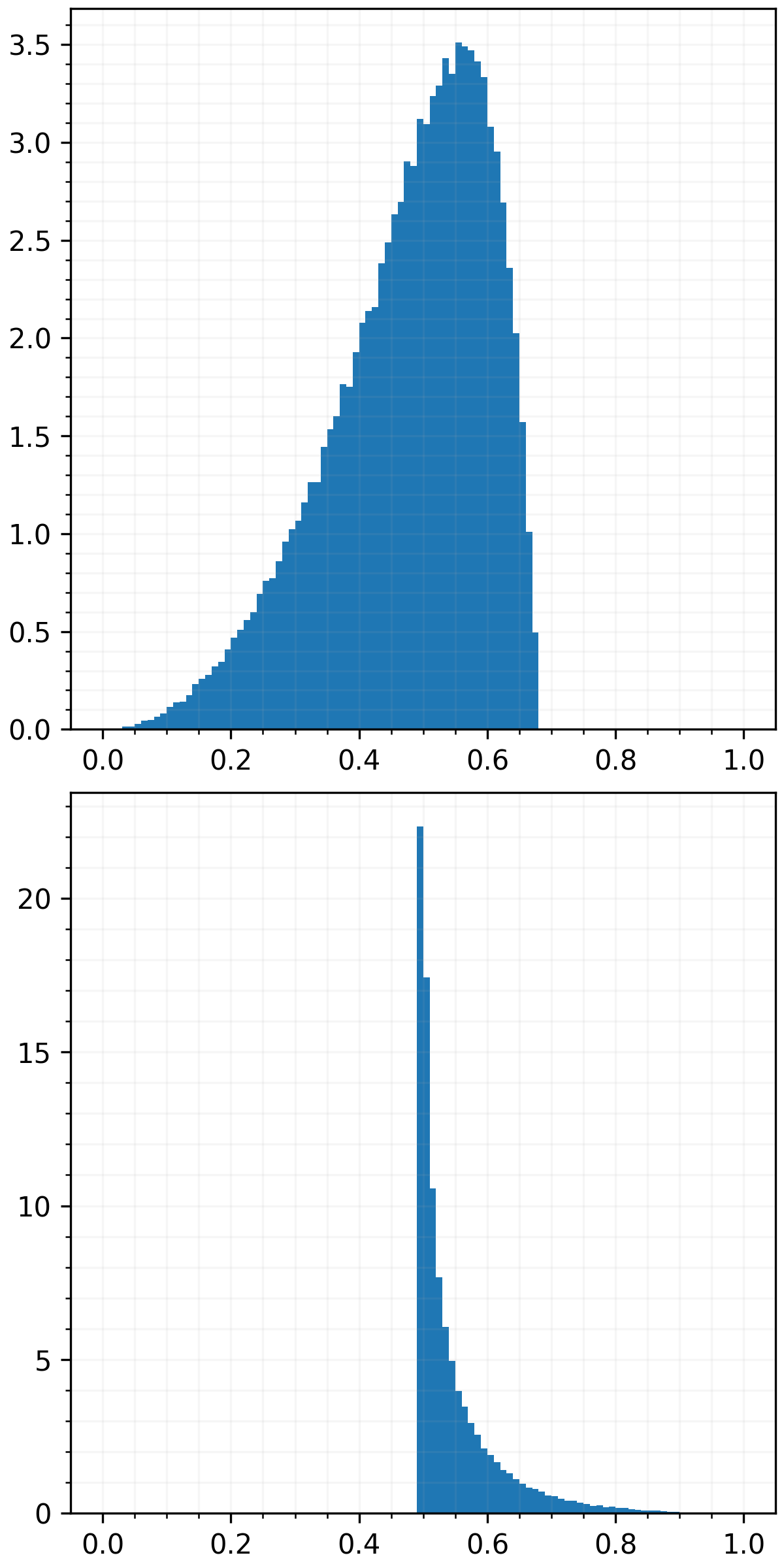}\hfil
    \includegraphics[width=.25\columnwidth]{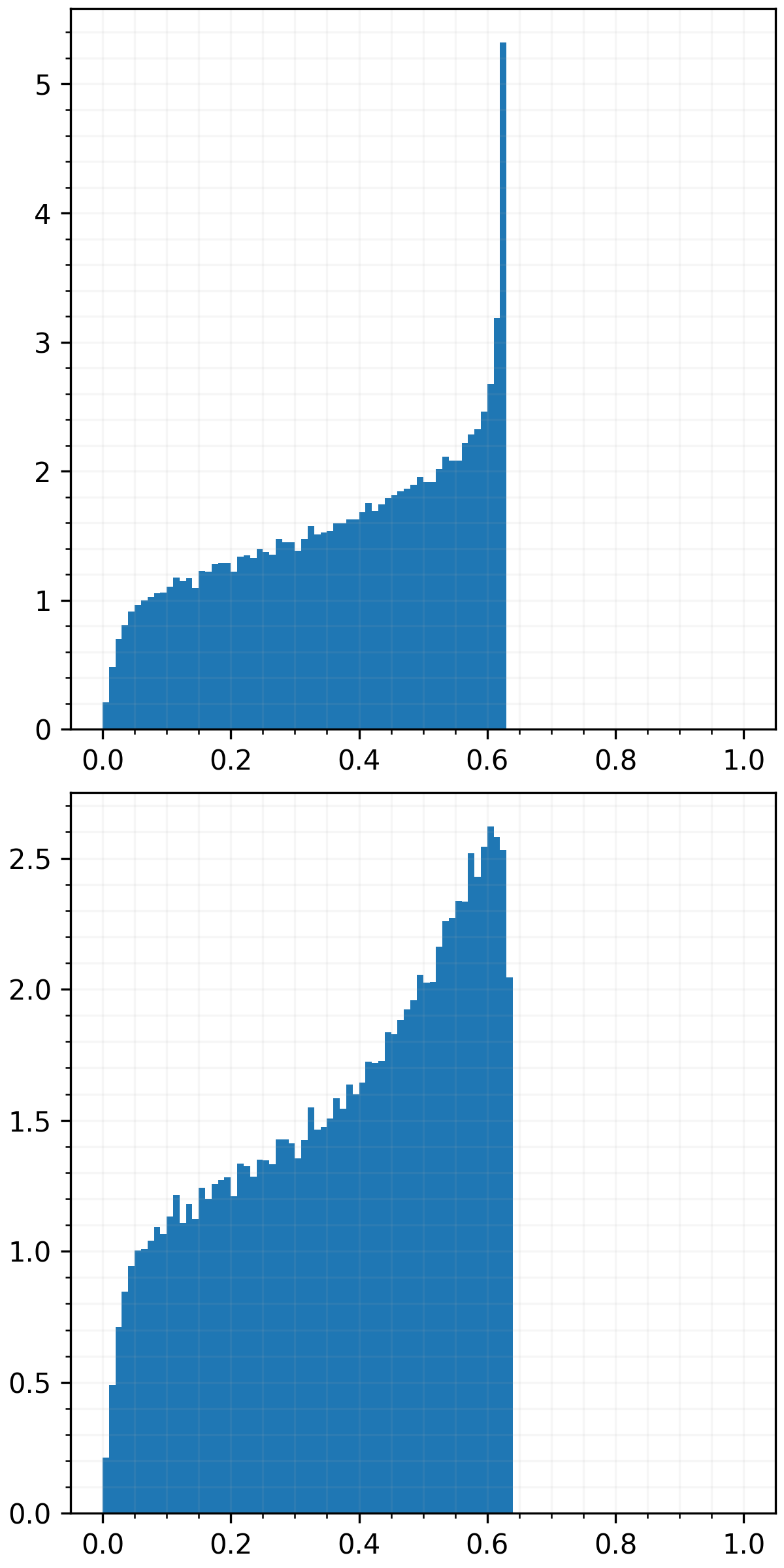}\hfil
    \includegraphics[width=.25\columnwidth]{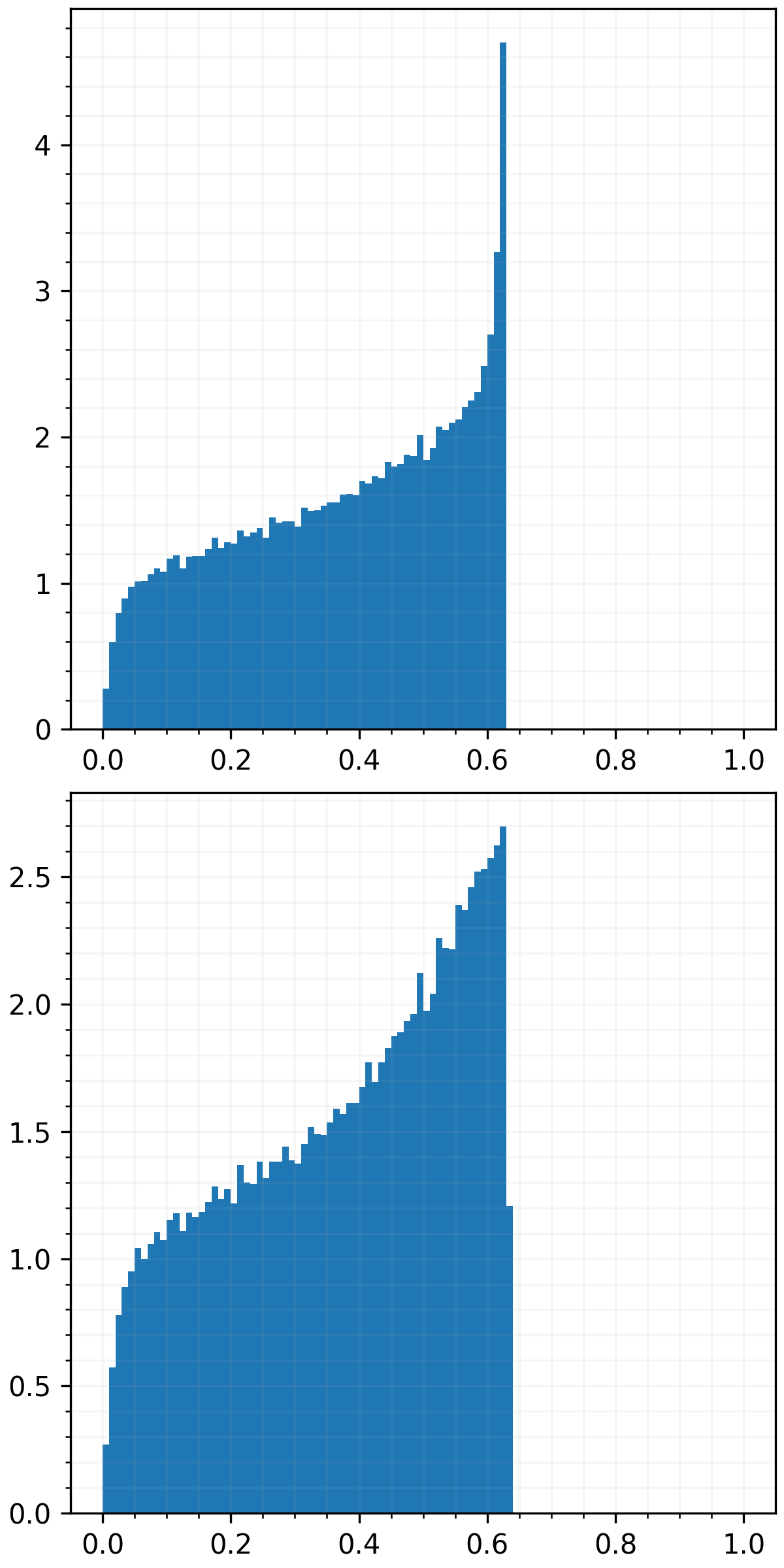}\hfil
    \includegraphics[width=.25\columnwidth]{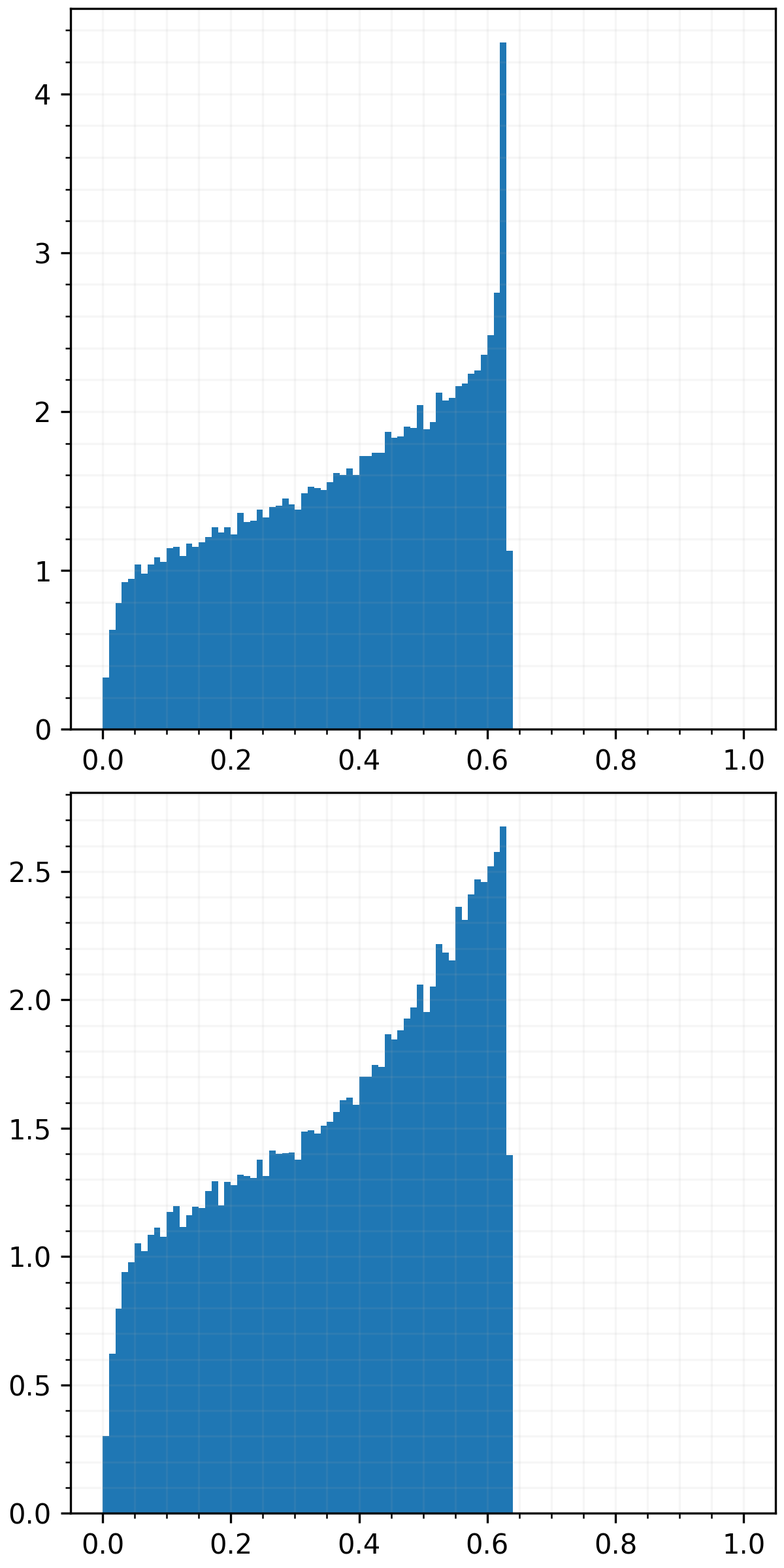}
    \caption{Strategies at epochs 0, 100, 200, and 300 (left to right). X and Y axes denote observation and bid, respectively. Each histogram uses \(10^5\) action samples.}
    \label{fig:visibility_strategies}
\end{figure}

Figures~\ref{fig:visibility_nashconv} illustrates performances on the 2-player visibility game. Figure~\ref{fig:visibility_strategies} illustrates strategies during training for a trial with 1-dimensional noise. The players' distributions converge to the expected distribution (there is a distinctive cutoff at \(1 - 1/\mathrm{e} \approx 0.632\)).
As expected, 0-dimensional noise, which yields deterministic strategies, performs very poorly. More interestingly, there is a noticeable gap in performance between 1-dimensional noise, which matches the dimensionality of the action space, and higher-dimensional noise. That is, using noise of higher dimension than the action space accelerates convergence in this game.

\section{Conclusions and future research}

We presented the first paper, to our knowledge, to solve general continuous-action games with unrestricted mixed strategies and without any gradient information. We accomplished this using zeroth-order optimization techniques that combine smoothed gradient estimators with equilibrium-finding gradient dynamics.
We modeled players' strategies using \emph{randomized policy networks} that take noise as input and can flexibly represent arbitrary observation-dependent, continuous-action distributions. Being able to model such mixed strategies is crucial for tackling continuous-action games that can lack pure-strategy equilibria.

We evaluated our method on various games, including continuous Colonel Blotto games, single-item and multi-item auctions, and a visibility game.
The experiments showed that our method can quickly compute high-quality approximate equilibria for these games. Furthermore, they show that the dimensionality of the input noise is crucial for representing and converging to equilibria. In particular, noise of too low dimension (or no noise, which yields a deterministic policy) results in failure to converge.
Randomized policy networks flexibly model observation-dependent action distributions.
Thus, in contrast to prior work, we can flexibly model mixed strategies and directly optimize them in a ``black-box'' game with access only to payoffs.

This work opens many directions for tackling even more complex multiagent environments.
In multi-step environments, the current observation may not contain all information about past observations and actions that is relevant to choosing an action.
To give agents memory, one can use recurrent networks~\citep{Rumelhart_1986, Werbos_1988} such as LSTMs~\citep{Hochreiter_1997} or GRUs~\citep{Cho_2014}.
In that case, the policy network would receive as input an observation, source of randomness, and memory state and output an action and new memory state.
One can also consider games with more complex observation and action spaces, including high-dimensional arrays like images. Convolutional networks~\citep{LeCun_1988, LeCun_1989} can be used to process such inputs. Very complex environments, including real-time strategy games like StarCraft II, may require more sophisticated neural architectures~\citep{Vinyals19:Grandmaster} such as pointer networks~\citep{Vinyals_2015}, transformers~\citep{Vaswani_2017}, and scatter connections that integrate spatial and non-spatial information.

\section{Acknowledgements}

This material is based on work supported by the National Science Foundation under grants IIS-1901403 and CCF-1733556 and by the ARO under award W911NF2210266.

\bibliography{dairefs,references}
\appendix
\section{Best response computation for continuous Colonel Blotto game}
Approximate best responses can be computed for the continuous Colonel Blotto game without discretizing the action space~\citep{Ganzfried_2021}.
By sampling \(K\) batches of actions from other players' strategies, we can obtain an approximate best response \(a_i\) for player \(i\) using a mixed-integer linear program (MILP). More precisely, let \(h_{jk}\) be the highest bid for \(j\) from other players in batch \(k\). Solve the following MILP:
\begin{align}
    \text{maximize} \quad
    & \textstyle\sum_j v_{ij} \tfrac{1}{K} \textstyle\sum_k z_{jk} \\
    \text{over} \quad
    & a_i \in \mathbb{R}^J \\
    & z \in \{0, 1\}^{J \times K} \\
    \text{subject to} \quad
    & a_{ij} \geq 0 \quad \forall j \\
    & \textstyle\sum_j a_{ij} = b_i \\
    & z_{jk} = [a_{ij} \geq h_{jk}] \quad \forall j, k
\end{align}
We can use the Big M method to represent the last constraint:
\begin{align}
    a_{ij} - h_{jk} + M(1 - z_{jk}) \geq 0
\end{align}
where \(M \gg 0\), which forces \(z_{jk}\) to be 0 when \(a_{ij} - h_{jk}\) is negative.

\section{Additional information about auctions}

Table~\ref{tab:value_structures} describes the independent private values, common values, affiliated values, complete information, and asymmetric information auctions, in that order.

\begin{table}[!h]
    \centering
    \begin{tabular}{|l|l|l|l|}
    \hline
    \(\Omega\) & \(\tau_i(\omega)\) & \(v_i(\omega)\) \\
    \hline
    \([0, 1]^n\) & \(\omega_i\) & \(\omega_i\) \\
    \([0, 1]^{n+1}\) & \(\omega_i \omega_{n+1}\) & \(\omega_{n+1}\) \\
    \([0, 1]^{n+1}\) & \(\omega_i + \omega_{n+1}\) & \(\omega_{n+1} + \tfrac{1}{n} \textstyle\sum_{i=1}^n \omega_i\) \\
    \([0, 1]\) & \(\omega\) & \(\omega\) \\
    \([0, 1]\) & \(\omega [i = 1]\) & \(\omega\) \\
    \hline
    \end{tabular}
    \caption{Auction descriptions. All use \(\mu = \mathcal{U}(\Omega)\).}
    \label{tab:value_structures}
\end{table}

For the common values auction, also known as the ``mineral rights'' model~\citep[example 6.1]{Krishna02:Auction}, the following procedure can be used to sample \(\omega \mid o_i\):
\begin{align}
    z &\sim \mathcal{U}([0, 1]) \\
    \omega_{n+1} &= o_i^z \\
    \omega_i &= o_i \mathbin{/} \omega_{n+1} \\
    \omega_j &\sim \mathcal{U}([0, 1]) \qquad j \neq i
\end{align}
For the affiliated values auction~\citep[example 6.2]{Krishna02:Auction}, the following procedure can be used to sample \(\omega \mid o_i\):
\begin{align}
    \omega_{n+1} &\sim \mathcal{U}(\max\{0,o_i-1\}, \min\{1,o_i\}) \\
    \omega_i &= o_i - \omega_{n+1} \\
    \omega_j &\sim \mathcal{U}([0, 1]) \qquad j \neq i
\end{align}

The all-pay auction with independent private values has a pure symmetric equilibrium generated by:
\begin{align}
    a_i &= \tfrac{n-1}{n} (o_i)^n
\end{align}
The \(k\)th-price winner-pay auction with independent private values has a pure symmetric equilibrium generated by~\citep[p. 878]{Kagel_1993}:
\begin{align}
    a_i &= \tfrac{n-1}{n+1-k} o_i
\end{align}
The 3-player 2nd-price winner-pay auction with common values has a pure symmetric equilibrium generated by~\citep{Bichler_2021}:
\begin{align}
    a_i &= \tfrac{o_i}{2 + \tfrac{1}{2} o_i}
\end{align}
The 2-player 1st- and 2nd-price winner-pay auction with affiliated values have pure symmetric equilibria generated by, respectively~\citep{Bichler_2021}:
\begin{align}
    a_i &= \tfrac{2}{3} o_i \\
    a_i &= o_i
\end{align}

\section{Other equilibrium-finding dynamics}
We tested a broad set of equilibrium-finding dynamics. This section describes those dynamics.
Let \(\xi\) denote the \emph{simultaneous gradient} of the utilities with respect to the parameters of the respective players only: \(\xi_i = \nabla_i u_i\).
This is a vector field on the strategy profile space. In general, it is not conservative (the gradient of some potential). This is the primary source of difficulties in applying standard gradient-based optimization methods, since trajectories can cycle around fixed points rather than converge to them. Various algorithms have been proposed in the literature to address this. These include the following:
\begin{enumerate}
\item Simultaneous gradient:
\begin{align}
    x^{t+1} = x^t + \alpha \xi(x^t)
\end{align}
\item Extragradient~\citep{Korpelevich_1976}:
\begin{align}
    x^{t+1} = x^t + \alpha \xi(x^t + \beta \xi(x^t))
\end{align}
\item Optimistic gradient~\citep{Daskalakis_2017, Mokhtari_2020}:
\begin{align}
    x^{t+1} = x^t + \alpha \xi(x^t) + \beta (\xi(x^t) - \xi (x^{t-1}))
\end{align}
\item \emph{Consensus optimization (CO)}~\citep{Mescheder_2017}:
\begin{align}
    x^{t+1} &= x^t + \alpha ( \xi - \beta \nabla \tfrac{1}{2} \|\xi\|^2 )(x^t) \\
    &= x^t + \alpha ( \xi - \beta \xi \cdot \nabla \xi)(x^t)
\end{align}
\item \emph{Symplectic gradient adjustment (SGA)}~\citep{Balduzzi_2018, Gemp_2018, Letcher_2019}:
\begin{align}
    x^{t+1} &= x^t + \alpha ( \xi - \beta \xi \cdot (\nabla \xi)_\mathrm{A} )(x^t)
\end{align}
\end{enumerate}
where \(M_\mathrm{A} = \tfrac{1}{2}(M - M^\top)\) is the antisymmetric part of \(M\). CO and SGA are second-order methods, meaning they require access to the second derivatives of the utility functions. CO can converge to bad critical points even in simple cases where the `game' is to minimize a single function.
Thus it cannot be considered a candidate algorithm for finding stable fixed points in general games, since it fails in the basic case of potential games~\citep[p. 13]{Letcher_2019}.
SGA was created to address some of the shortcomings of CO.

\section{Analytically-derived equilibria of the benchmark games}

We carefully selected the benchmark games to be ones for which an analytically-derived equilibrium exists, so that we can compare the computationally-derived equilibria to the analytical one. In this section, we illustrate the analytical equilibria to support that comparison.

\begin{figure}[!h]
    \centering
    \includegraphics[width=.8\columnwidth]{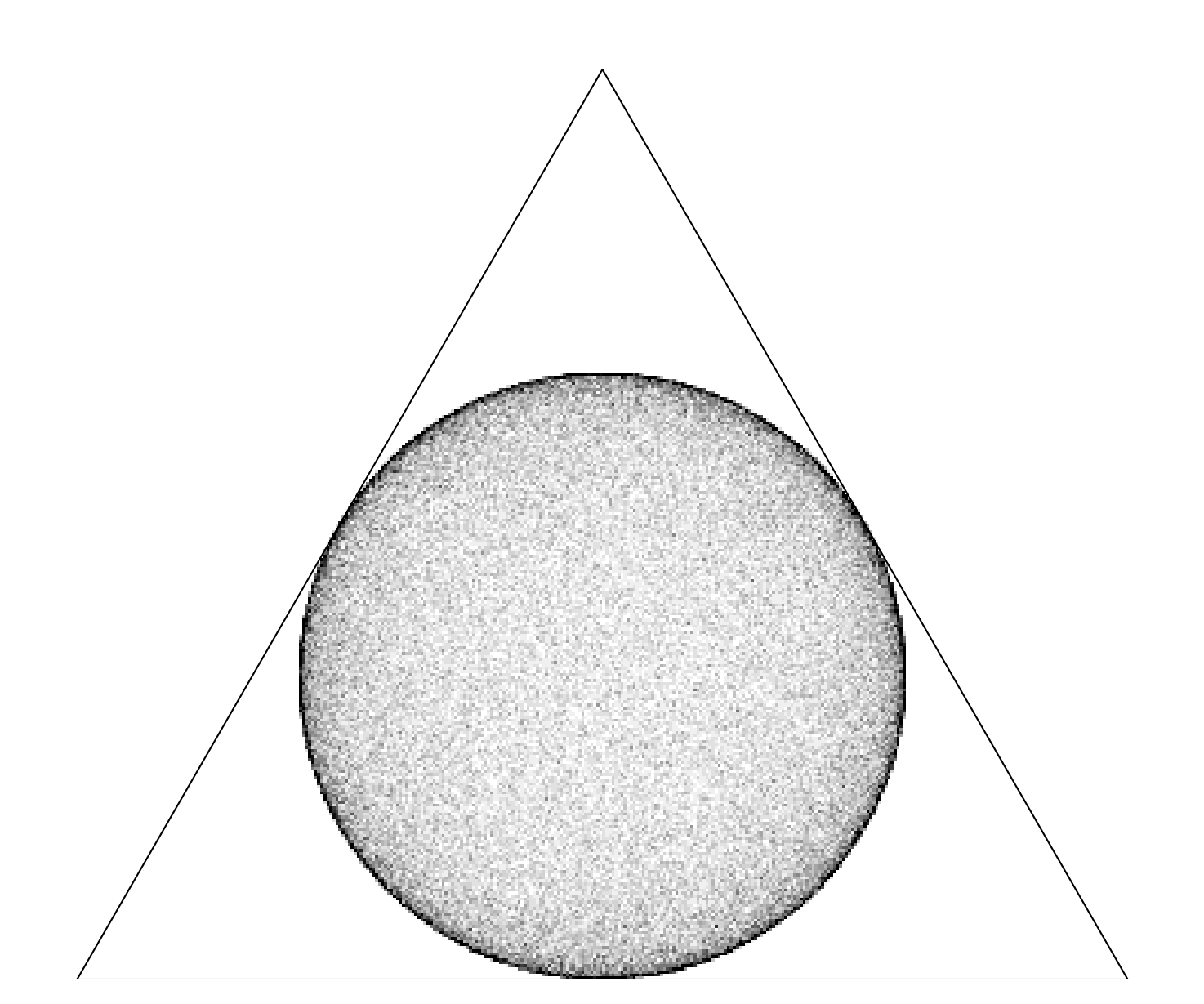}
    \caption{Continuous Colonel Blotto game solution.}
    \label{fig:solution_blotto}
\end{figure}

Figure~\ref{fig:solution_blotto} illustrates the analytical solution for the continuous Colonel Blotto game with fixed homogeneous budgets and valuations. This game was analyzed by~\citet{Gross_1950}, who also analyzed the game for various special cases of heterogeneous budgets and valuations. They give the following geometric description of the equilibrium strategy: ``[The player] inscribes a circle within [the triangle] and erects a hemisphere upon this circle. He next chooses a point from a density uniformly distributed over the surface of the hemisphere and projects this point straight down into the plane of the triangle... He then divides his forces in respective proportion to the triangular areas subtended by [this point] and the sides.''

\begin{figure}
    \centering
    \includegraphics[width=.8\columnwidth]{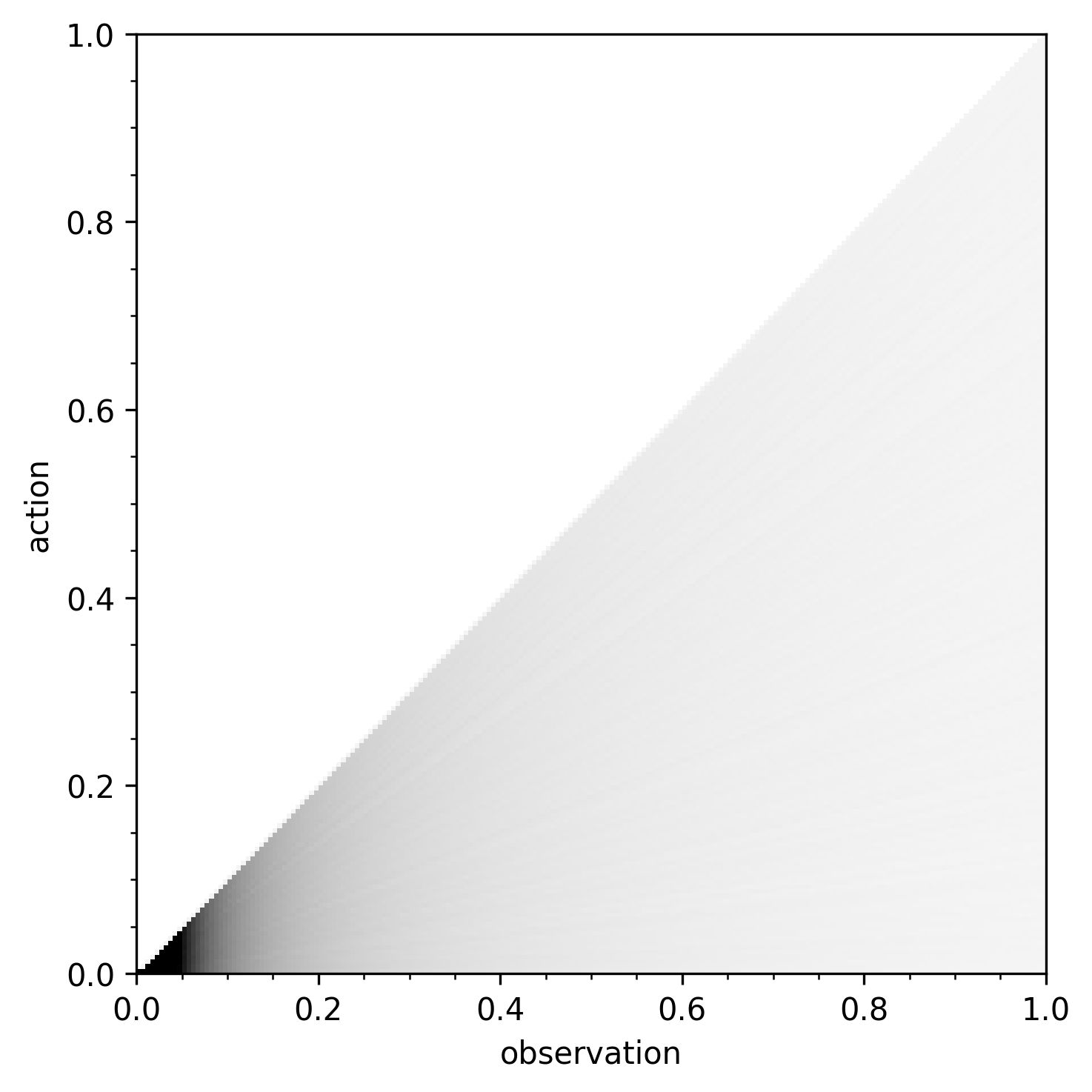}
    \caption{Complete-information auction solution.}
    \label{fig:solution_complete}
\end{figure}

Figure~\ref{fig:solution_complete} illustrates the analytical solution for the all-pay complete information auction. This game was analyzed by~\citet{Baye_1996}. They show that with homogeneous valuations (\(v_1 = v_2 = \ldots = v_n\)) there exists a unique symmetric equilibrium and a continuum of asymmetric equilibria. All of these equilibria are payoff equivalent, as is the expected sum of the bids (revenue to the auctioneer). In the symmetric equilibrium, each player randomizes uniformly on \([0, v_1]\).

\begin{figure}
    \centering
    \includegraphics[width=.8\columnwidth]{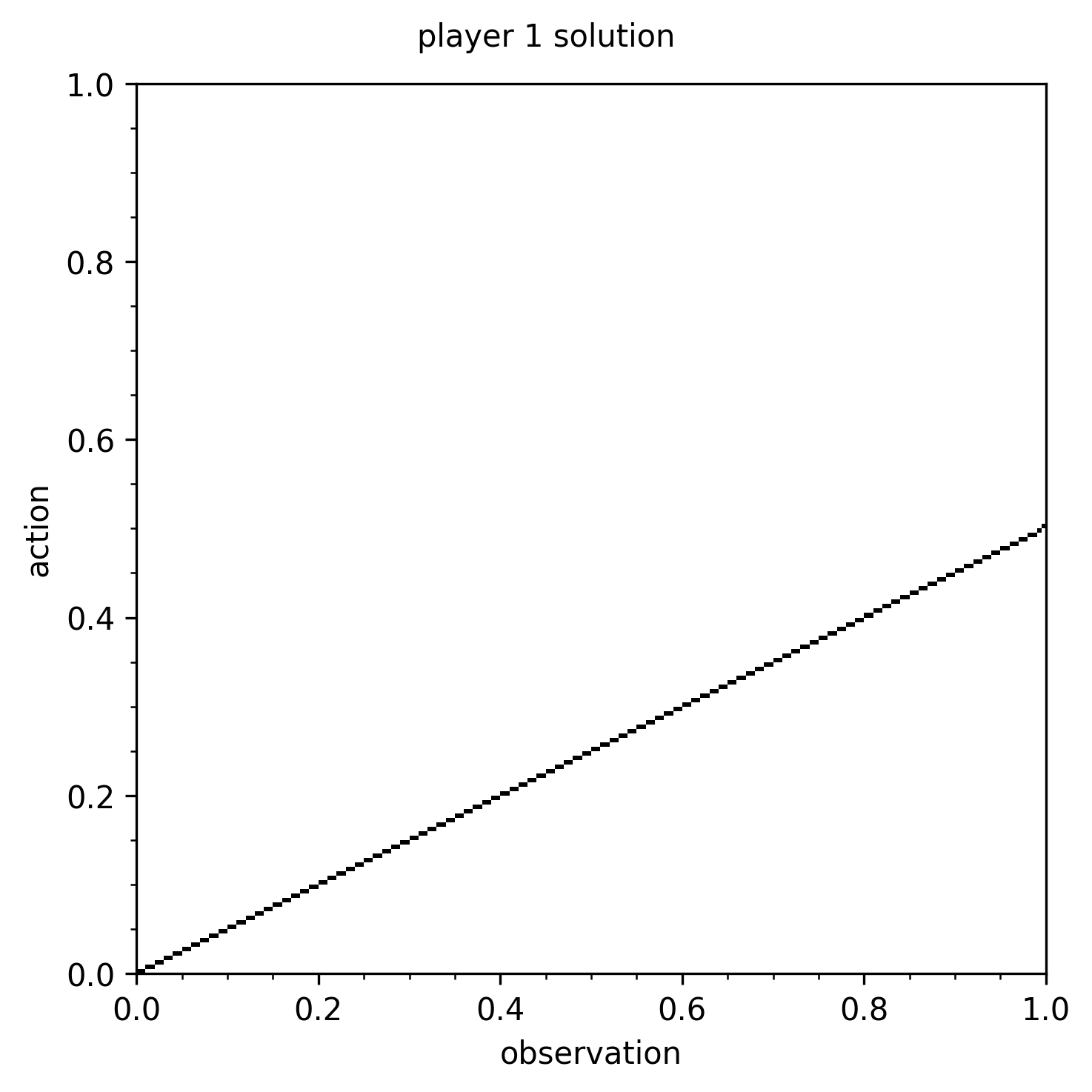}
    \includegraphics[width=.8\columnwidth]{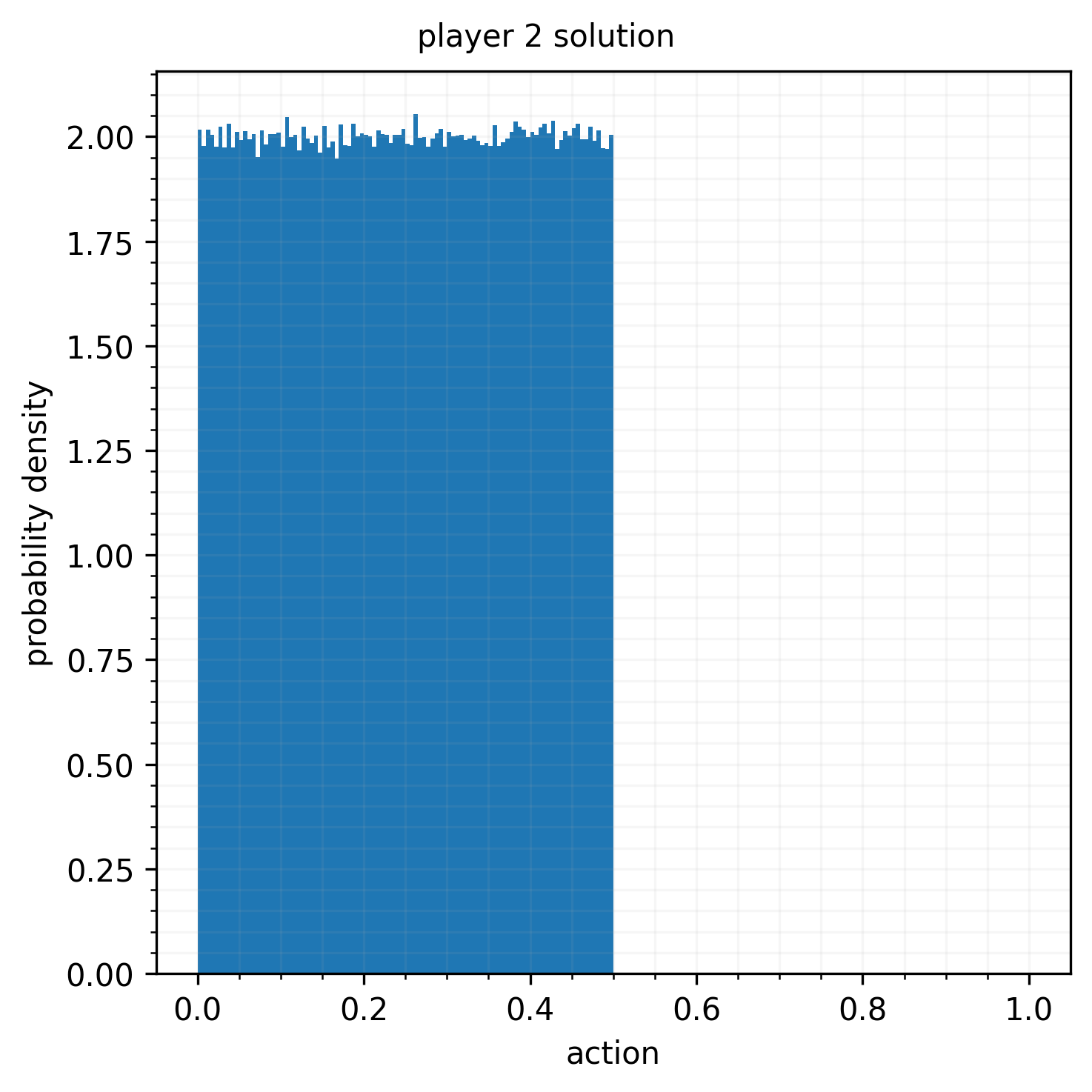}
    \caption{Asymmetric-information auction solution.}
    \label{fig:solution_asymmetric}
\end{figure}

Figure~\ref{fig:solution_asymmetric} illustrates the analytical solution for the 2-player 1st-price winner-pay asymmetric-information auction. This game was analyzed by~\citet[section 8.3]{Krishna02:Auction}.
Bidder 1 bids according to the strategy \(\beta(v) = \operatorname{E}[V \mid V \leq v]\). In our case, \(V \sim \mathcal{U}([0, 1])\), so \(\beta(v) = \tfrac{v}{2}\).
Bidder 2 chooses a bid at random from the interval \([0, \operatorname{E}[V]]\) according to the distribution defined by \(H(b) = \operatorname{P}[\beta(V) \leq b]\). In our case, this reduces to the distribution \(\mathcal{U}([0, \tfrac{1}{2}])\).

\begin{figure}[!h]
    \centering
    \includegraphics[width=.8\columnwidth]{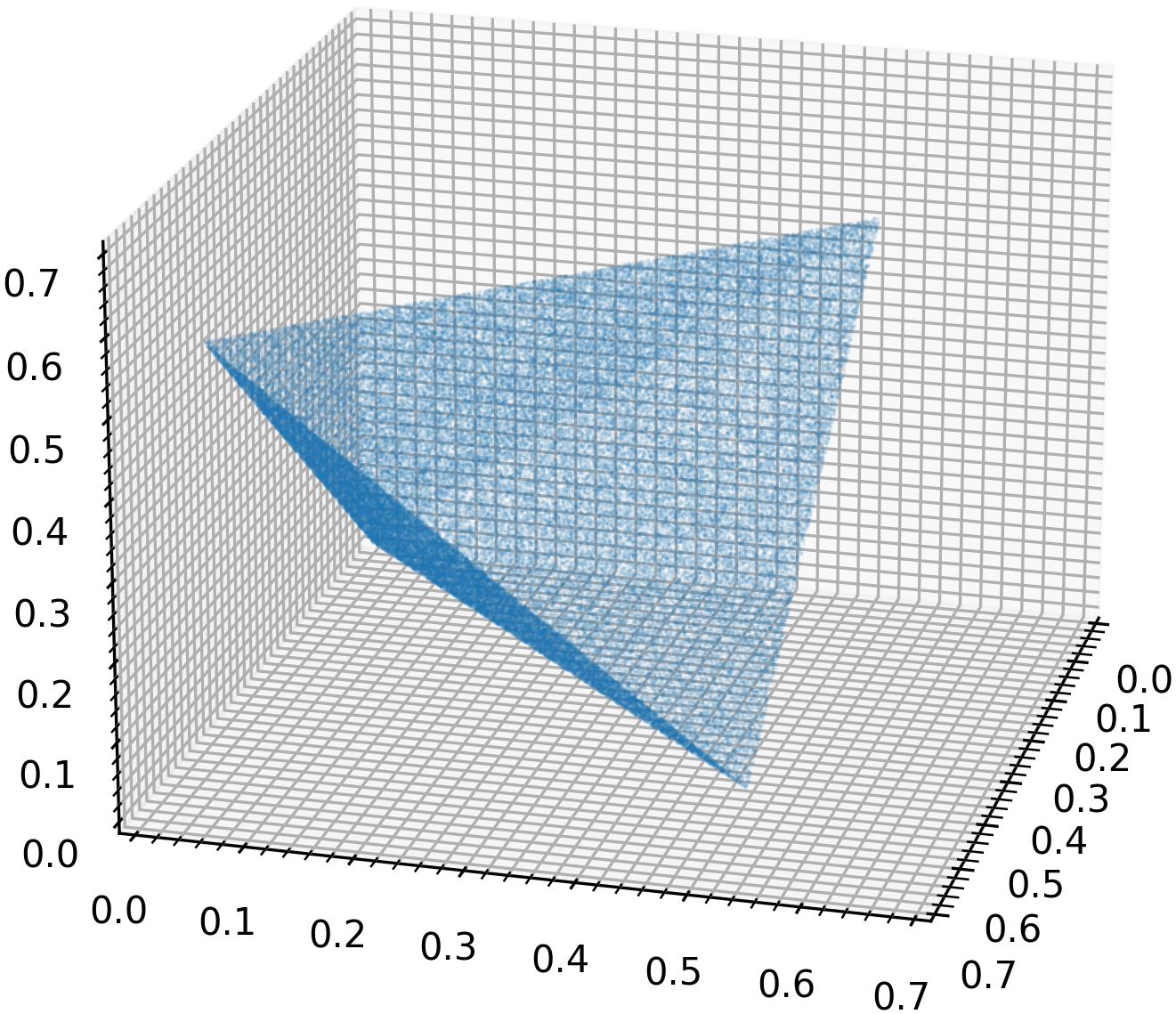}
    \caption{Chopstick auction solution.}
    \label{fig:solution_chopstick}
\end{figure}

Figure~\ref{fig:solution_chopstick} illustrates the analytical solution for the chopstick auction. This game was analyzed by~\citet{Szentes_2003, Szentes03:Chopsticks}, who state the following: ``The supports of the mixtures that generate the symmetric equilibria in both the first- and second-price cases, turn out to be the surfaces of regular tetrahedra, and the distributions themselves turn out to be uniform on these surfaces. In addition, in each case all the points inside the tetrahedron are pure best responses to the equilibrium mixture.''

\begin{figure}[!h]
    \centering
    \includegraphics[width=.8\columnwidth]{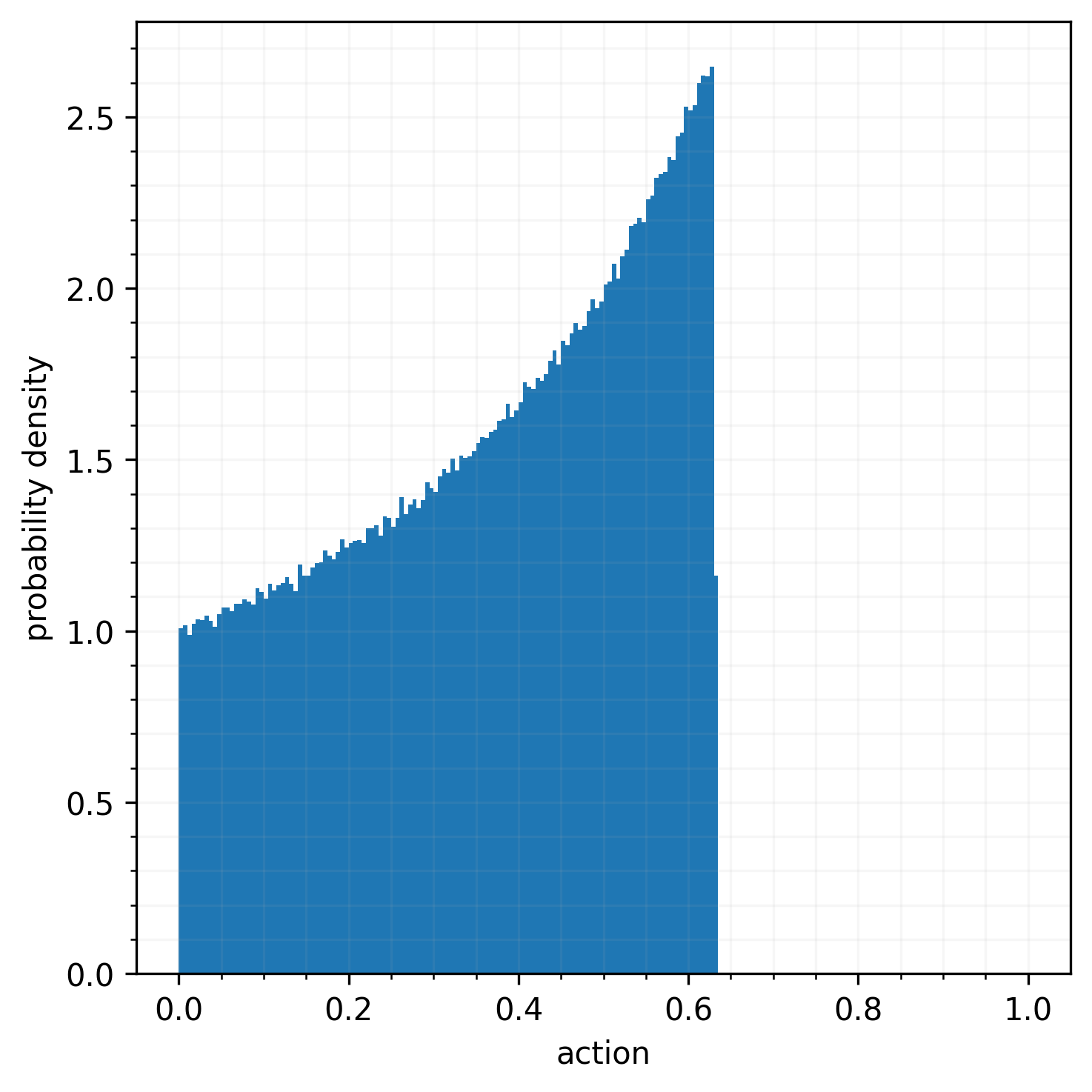}
    \caption{Visibility game solution.}
    \label{fig:solution_visibility}
\end{figure}

Figure~\ref{fig:solution_visibility} illustrates the analytical solution for the 2-player visibility game. This game was analyzed by~\citet{Lotker_2008}. Up to a set of measure zero, it has a unique equilibrium whose strategies have probability densities \(p(x) = 1/(1-x)\) when \(0 \leq x \leq 1 - 1/\mathrm{e}\) and 0 otherwise. Each player's expected payoff is \(1/\mathrm{e}\).

\end{document}